\documentclass[10pt,journal]{IEEEtran}
\usepackage{cite,comment}

\ifCLASSINFOpdf
\else
\fi

\usepackage{amsmath,graphicx}
\usepackage{amssymb,amsfonts}
\usepackage{booktabs}
\usepackage{multirow}
\usepackage[english]{babel}
\usepackage{amsthm}
\usepackage{algorithm,algorithmic}
\usepackage{caption}
\usepackage{array}
\usepackage{slashbox}
\usepackage{mathrsfs}
\usepackage{bbm}
\usepackage{xcolor}
\usepackage{colortbl}
\allowdisplaybreaks

\def\Mat#1{\boldsymbol{\mathbf{#1}}}

\def\Vec#1{\textsf{\boldmath $#1$}}
\def\ceq{\overset{c}{=}}
\def\E{{\mathbb E}}
\def\C{{\mathbb C}}
\def\T{{\textsf{T}}}
\def\H{{\textsf{H}}}
\def\Normal{{\mathcal{N}}}

\def\Diag{\mathop{\mathrm{diag}}\nolimits}
\def\Argmax{\mathop{\rm{argmax}}}

\newcommand{\refeq}[1]{(\ref{eq:#1})}
\newcommand{\refeqs}[2]{(\ref{eq:#1}) and (\ref{eq:#2})}

\newcommand{\refsec}[1]{Section \ref{sec:#1}}
\newcommand{\refsubsec}[1]{Subsection \ref{subsec:#1}}

\newcommand{\reffig}[1]{Fig. \ref{fig:#1}}

\newcommand{\reftab}[1]{Table \ref{tab:#1}}

\newcolumntype{C}[1]{>{\centering\arraybackslash\hspace{0pt}}m{#1}}

\begin{document}
\title{FastMVAE2: On improving and accelerating \\
the fast variational autoencoder-based \\
source separation algorithm for determined mixtures
}

\author{Li~Li,~\IEEEmembership{Member,~IEEE,}
	  Hirokazu~Kameoka,~\IEEEmembership{Senior Member,~IEEE,}
      and Shoji Makino,~\IEEEmembership{Fellow,~IEEE,}
\thanks{
L. Li and H. Kameoka are with NTT Communication Science Laboratories, NTT Corporation, 3-1 Morinosato Wakamiya, Atsugi-shi, Kanagawa 243-0198, Japan. L. Li is currently also with Nagoya University, Furo-cho, Chikusa-ku, Nagoya, 464-8601, JAPAN, (email: lili-0805@ieee.org, hirokazu.kameoka.uh@hco.ntt.co.jp).

S. Makino is with Waseda University, 2-7 Hibikino, Wakamatsu-ku, Kitakyushu city, Fukuoka, 808-0135, Japan, and University of Tsukuba, 1-1-1 Tennodai, Tsukuba, Ibaraki, 2050821, Japan, (email: s.makino@waseda.jp).

This work was partially supported by JST CREST JPMJCR19A3. A preprint version of this paper has already been made publicly available at \cite{arxiv}.
}
}

\markboth{Journal of \LaTeX\ Class Files,~Vol.~6, No.~1, January~2007}
{Shell \MakeLowercase{\textit{et al.}}: Bare Demo of IEEEtran.cls for Journals}

\maketitle

\begin{abstract}
This paper proposes a new source model and training scheme to improve the accuracy and speed of the multichannel variational autoencoder (MVAE) method. 
The MVAE method is a recently proposed powerful multichannel source separation method. It consists of pretraining a source model represented by a conditional VAE (CVAE) and then estimating separation matrices along with other unknown parameters so that the log-likelihood is non-decreasing given an observed mixture signal. Although the MVAE method has been shown to provide high source separation performance, one drawback is the computational cost of the backpropagation steps in the separation-matrix estimation algorithm. 
To overcome this drawback, a method called ``FastMVAE'' was subsequently proposed, which uses an auxiliary classifier VAE (ACVAE) to train the source model. 
By using the classifier and encoder trained in this way, the optimal parameters of the source model can be inferred efficiently, albeit approximately, in each step of the algorithm.
However, the generalization capability of the trained ACVAE source model was not satisfactory, which led to poor performance in situations with unseen data.
To improve the generalization capability, this paper proposes a new model architecture (called the ``ChimeraACVAE'' model) and a training scheme based on knowledge distillation. 
The experimental results revealed that the proposed source model trained with the proposed loss function achieved better source separation performance with less computation time than FastMVAE.
We also confirmed that our methods were able to separate 18 sources with a reasonably good accuracy.
\end{abstract}

\begin{IEEEkeywords}
Multichannel source separation, multichannel variational autoencoder (MVAE), fast algorithm, auxiliary classifier VAE, knowledge distillation
\end{IEEEkeywords}

\IEEEpeerreviewmaketitle

\section{Introduction}
\label{sec:introduction}
\IEEEPARstart{B}{lind} source separation (BSS) is a technique for separating observed signals recorded by a microphone array into individual source signals without prior information about the sources or mixing conditions.
This technique has been used in a wide range of applications, including hearing aids, automatic speech recognition (ASR), telecommunications systems, music editing, and music information retrieval.
\par
Acoustic signals are convolved with the impulse responses of acoustic environments and so the signal observed at a particular position is usually given as the convolutive mixture of nearby source signals. 
Although it is possible to take a time-domain approach to the BSS problem, it can be computationally expensive since it requires directly estimating and applying demixing filters with thousands of taps. 
In contrast, the time-frequency-domain approach is advantageous in that the convolution operations can be replaced by multiplications to achieve computationally efficient algorithms, and it allows the flexible use of various models for the time-frequency (TF) representations of source signals.
Independent vector analysis (IVA) \cite{Kim2006independent,Hiroe2006solution} is an example of the time-frequency-domain approach, which makes it possible to solve frequency-wise source separation and permutation alignment simultaneously by assuming that the magnitudes of the frequency components originating from the same source vary coherently over time.
Multichannel nonnegative matrix factorization (MNMF)  \cite{Ozerov2010multichannel,Sawada2013multichannel} and independent low-rank matrix analysis (ILRMA)  \cite{Kameoka2010statistical,Kitamura2016determined,Kitamura2018determined} are other examples, which employ the concept of NMF \cite{Lee2001algorithms} to model the TF structures of sources. 
Specifically, they assume that the power spectrum of each source signal can be approximated as the sum of a limited number of basis spectra scaled by time-varying amplitudes. 
IVA can be understood as a special case of ILRMA where only one flat basis spectrum is used for representing each source. 
This indicates that ILRMA can capture the TF structure of each source more flexibly than IVA, and this flexibility has been shown to be advantageous in improving the source separation performance \cite{Kitamura2016determined}. 
\par
Recently, the success of deep neural network (DNN)-based speech separation methods \cite{Wang2018supervised,Hershey2016deep,Isik2016single,Yu2017permutation,Kol2017multitalker,Liu2019divide,Roux2019phasebook,Delfarah2019deep,Wang2018multi}, including deep clustering (DC) \cite{Hershey2016deep,Isik2016single} and permutation invariant training (PIT) \cite{Yu2017permutation,Kol2017multitalker}, has proven that DNNs have an excellent ability to capture and learn the structure of spectrograms.
The general idea of these methods is to train a network that predicts a TF mask or clean signals given the spectral and spatial features of observed mixture signals. 
Meanwhile, time-domain methods based on end-to-end training have also been extensively studied and have shown excellent performance \cite{Luo2019conv,Nachmani2020voice,Subakan2021attention}. 
Some attempts \cite{Higuchi2017deep,Ochiai2020beam} have been made to combine beamforming with the time-domain methods to avoid artifacts introduced by nonlinear processing. 
Although such an end-to-end approach provides reasonably good separation performance, one drawback is that it suffers from the limitation that the test conditions need to be similar to the training ones, such as the number of speakers and reverberation conditions.
\par
There have also been some attempts to incorporate DNNs into the BSS methods mentioned earlier \cite{Nugraha2016multichannel,Makishima2019independent,Kameoka2019supervised,Bando2018statistical,Leglaive2018a,Li2020determined}.
Independent deeply low-rank matrix analysis (IDLMA) \cite{Mogami2018independent,Makishima2019independent} is one such method, where each DNN is trained using the utterances of a different speaker. After training, the trained DNNs are used to refine the estimated power spectra at each iteration of the source separation algorithm. Namely, each DNN can be seen as a speaker-dependent speech enhancement system.
One drawback of IDLMA would be that it can perform poorly in speaker-independent scenarios due to its discriminative training scheme.
Within the DNN framework, deep generative models such as variational autoencoders (VAEs) \cite{Kingma2014semi,Sohn2015learning}, generative adversarial networks (GANs) \cite{Goodfellow2014generative},  and normalizing flow (NF) \cite{Rezende2015variational} have proven to be powerful in source separation tasks \cite{Kameoka2019supervised,Bando2018statistical,Leglaive2018a,Sekiguchi2018bayesian,Sekiguchi2019semi,Leglaive2019speech,Leglaive2019semi,Seki2019underdetermined,Inoue2019joint,Li2020determined,Nugraha2020flow,Neri2021unsupervised,Bando2021neural}.
An attempt to employ VAE for semi-supervised single-channel speech enhancement was made in  \cite{Bando2018statistical} under the name of the ``VAE-NMF'' method, which uses a VAE to model each single-frame spectrum in an utterance of a target speaker and an NMF model to express a noise spectrogram. Several variants of this method have subsequently been developed, including the incorporation of loudness gain for robust speech modeling \cite{Leglaive2018a}, the adoption of a noise model based on alpha-stable distribution instead of a complex Gaussian distribution \cite{Leglaive2019speech}, and the extension to multichannel scenarios \cite{Sekiguchi2019semi,Leglaive2019semi}. 
\par
Independently, around the same time, we proposed a method called the ``multichannel variational autoencoder (MVAE)''. This was the first to incorporate the VAE concept into the multichannel source separation framework, and it has proven to be very successful in supervised determined source separation tasks. 
Unlike the VAE-NMF methods, the MVAE method uses a conditional VAE (CVAE) with a fully convolutional architecture to model the entire spectrogram of each utterance.
The CVAE is trained with the spectrograms of clean speech samples along with the corresponding speaker ID as a conditioning class variable. This is done so that the trained decoder distribution can be used as a generative model of signals produced by all the sources included in a given training set, where the latent space variables and the class variables are the parameters to be estimated from an input mixture signal.
The generative model trained in this way is called the {\it CVAE source model}. 
At the separation phase, the MVAE algorithm iteratively updates the separation matrix using the iterative projection (IP) method \cite{Ono2011stable}
and the underlying parameters of the CVAE source model using a gradient descent method, where the gradients of the latent variables are calculated using backpropagation. 
The main feature of this optimization algorithm is that the log-likelihood is guaranteed to be non-decreasing if the step size is carefully chosen or if a backtracking line search \cite{Sun2006optimization} is applied for the backpropagation algorithm.
Furthermore, since the MVAE uses a CVAE to model single source and the demixing matrices are estimated only at separation phase, a trained CVAE source model is principle able to handle arbitrary number of sources and different recording conditions, which is significantly differ from discriminative methods.
However, one major drawback of the MVAE method is that the backpropagation required for each iteration makes the optimization algorithm very time-consuming, which can be problematic in practice. 
\par
To address this problem, we previously proposed a fast algorithm called ``FastMVAE" \cite{Li2020fast}, which uses an auxiliary classifier VAE (ACVAE) \cite{Kameoka2019acvae} 
to model the generative distribution of source spectrograms. 
In this method, the encoder and auxiliary classifier are trained in such a way that they learn to infer the latent space variables and class variables, respectively, given a spectrogram.
This allows us to replace the backpropagation steps in the source separation algorithm with the forward propagation of the two networks and thus significantly reduce the computational cost. 
Furthermore, we showed that FastMVAE can achieve source separation performance comparable to the MVAE method when the training and test conditions are sufficiently close to being consistent.
However, when there is mismatch between the training and test conditions, due to, for example, the presence of long reverberation or under speaker-independent conditions, FastMVAE tends to perform worse than the MVAE method. This may be because the encoder and classifier cannot generalize well to inputs that are very different from the training data. To stabilize the parameter inference process under such mismatched conditions, we derived an improved update rule based on the Product-of-Experts (PoE) framework \cite{Hinton2002training}. However, this method requires manual selection of the optimal weights in advance, forcing us to rely on heuristics.  
\par
FastMVAE being weak against the mismatch between the training and test conditions may be because the model is structured in such a way that the output of the auxiliary classifier is fed into the encoder and so the error in the classifier output can directly affect the encoder output. 
One way to avoid this would be to assume a conditional independence between the outputs of the encoder and auxiliary classifier so that they can perform their tasks in parallel.
Instead of preparing two separate networks, we propose merging the encoder and classifier into a single multitask network to allow them to share information. 
We call this new model the {\it ``ChimeraACVAE" source model}. 
\par
Another important issue is how to train the above model to have good generalization ability.
A number of techniques have been developed with the aim of improving the generalization ability of DNNs. 
These techniques can be roughly classified into regularization-based \cite{dropout,weightdecay,distillation,multitask}, data augmentation-based \cite{augmentation}, and training strategy-based methods \cite{transfer,batchnormalization,layernormalization}. 
Knowledge distillation (KD), a model compression and acceleration technique that has been rapidly gaining attention in recent years, is typically used to transfer knowledge of a teacher model to a more compact student model. KD has been shown to not only accelerate the inference process through model compression but also provide better generalization ability to the compressed model.
In this paper, we propose adopting KD to train the ChimeraACVAE source model. 
Specifically, we use a pretrained CVAE model as a teacher model and transfer its knowledge to the ChimeraACVAE model by using as a criterion the Kullback-Leibler (KL) divergence between the distributions of the outputs of the encoder and decoder of the CVAE and ChimeraACVAE models. 

In summary, the two main contributions of this paper are as follows:
\begin{itemize}
	\item We propose a new network architecture that replaces the ACVAE source model in FastMVAE, which we call the {\it ``ChimeraACVAE" source model}. It merges the encoder and classifier into a single multitask network so that it can handle the tasks of the encoder and classifier simultaneously.
	\item We propose a loss function based on the KD framework that allows the ChimeraACVAE source model to acquire excellent generalization capability. We show that the model trained in this way can improve source separation performance in both speaker-dependent and speaker-independent conditions.
\end{itemize}
\par
The rest of this paper is structured as follows.
After describing the formulation of the determined multichannel BSS problem and reviewing the original MVAE method in \refsec{MVAE}, we describe the ACVAE source model and the FastMVAE method in \refsec{FastMVAE}.
In \refsec{proposed}, we provide technical details of the proposed ChimeraACVAE source model and its training strategy.
The effectiveness of the proposed method is demonstrated in \refsec{experiment} by evaluating the source separation performance of speaker-dependent and speaker-independent scenarios. We conclude the article in \refsec{conclusion}.

\section{MVAE}
\label{sec:MVAE}
\subsection{Problem Formulation}
Let us consider a situation where $I$ source signals are captured by $I$ microphones. 
We use $x_i(f, n)$ and $s_j(f, n)$ to denote the short-time Fourier transform (STFT) coefficients of the signal observed at the $i$th microphone and $j$th source signal, 
where $f$ and $n$ are the frequency and time indices, respectively.
If we use 
\begin{align}   
	\Mat{x}(f, n) &= [x_1(f, n), \ldots, x_I(f, n)]^{\T} \in \C^I,  \\
	\Mat{s}(f, n) &= [s_1(f, n), \ldots, s_I(f, n)]^{\T} \in \C^I,
\end{align}
to denote 
the vectors containing $x_1(f, n), \ldots, x_I(f, n)$ and $s_1(f, n), \ldots, s_I(f, n)$,
the relationship between the observed signals and source signals can be approximated as
\begin{align}
	\Mat{s}(f, n) &= \Mat{W}^{\H}(f)\Mat{x}(f, n), 
	\label{eq:demixing} \\
	\Mat{W}(f) &= [\Mat{w}_1(f), \ldots, \Mat{w}_I(f)] \in \C^{I \times I},
\end{align}
under a determined mixing condition, 
where $\Mat{W}^{\H}(f)$ represents the separation matrix, and
$(\cdot)^{\T}$ and $(\cdot)^{\H}$ denote the transpose and Hermitian transpose of a matrix or a vector, respectively. 
The goal of BSS is to determine $\mathcal{W}=\{\Mat{W}(f)\}_f$ solely from the observation $\mathcal{X}=\{\Mat{x}(f, n)\}_{f,n}$.
Here, the notation $\{\Vec{E}_b\}_b$ is used as an abbreviation for $\{ \Vec{E}_b \mid b \in \mathcal{B} \}$, where $\mathcal{B}$ denotes the set of all possible indices. 
\par
In the following, we assume that 
$s_j(f, n)$ independently follows a zero-mean complex proper Gaussian distribution with variance (power spectral density) $v_j(f, n)=\E[|s_j(f, n)|^2]$:
\begin{align}
	p(s_j(f, n)|v_j(f,n))= \Normal_{\C} (s_j(f, n)|0,v_j(f,n)),
\label{eq:LGM}
\end{align}
This assumption is often referred to as the local Gaussian model (LGM) \cite{Fevotte2005maximum,Vincent2009underdetermined}.
If $s_j(f, n)$ and $s_{j'}(f, n)$ are independent for $\forall j \neq j'$, 
the density of $\Mat{s}(f,n)$
becomes
\begin{align}
	p(\Mat{s}(f, n)|\Mat{V}(f,n)) 
	&= \prod_j p(s_j(f, n)|v_j(f,n)) \nonumber\\
	&=
	\Normal_{\C}(\Mat{s}(f,n) | \Mat{0}, \Mat{V}(f, n)),
\label{eq:LGM2}
\end{align}
where $\Mat{V}(f, n)=\Diag[v_1(f, n), \ldots, v_I(f, n)]$. 
From \refeq{demixing} and \refeq{LGM2}, 
the density of $\Mat{x}(f,n)$ is obtained as
\begin{multline}
p(\Mat{x}(f,n)|\Mat{W}(f), \Mat{V}(f,n))=\\
|\Mat{W}^{\H}(f)|^{2} p(\Mat{s}(f,n)= \Mat{W}^{\mathsf H}(f)\Mat{x}(f,n)|\Mat{V}(f,n)),
\end{multline}
where $|\Mat{W}^{\H}(f)|^{2}$ is the Jacobian of the mapping $\Mat{x}(f,n)\mapsto \Mat{s}(f,n)$.
Therefore, the log-likelihood of $\mathcal{W}=\{\Mat{W}(f)\}_f$ and $\mathcal{V}=\{v_j(f,n)\}_{f,n,j}$, given $\mathcal{X}=\{\Mat{x}(f, n)\}_{f, n}$ is expressed as
\begin{align}
\label{eq:ll}
	&\log p(\mathcal{X}|\mathcal{W}, \mathcal{V}) \nonumber\\
	&= 2N \sum_f \log |\det \Mat{W}^{\H}(f)| 
	+ \sum_{j} \log p(\Vec{S}_j|\Vec{V}_j)\nonumber\\
	&\ceq 2N \sum_f \log |\det \Mat{W}^{\H}(f)| \nonumber\\
	&~~~~~~~~~~~~
	-\sum_{f, n, j} \bigg (\log v_j(f, n) + \frac{|\Mat{w}_j^{\H}(f) \Mat{x}(f, n)|^2}{v_j(f, n)} \bigg),
\end{align}
where we have used
$\ceq$ to denote equality up to constant terms 
and a bold italic font to indicate a set consisting of TF elements,
namely
$\Vec{S}_j = \{s_j(f,n)\}_{f,n}$ 
and $\Vec{V}_j = \{v_j(f,n)\}_{f,n}$.
The log-likelihood
will be split into $F$ frequency-wise terms if no additional constraint is imposed on $v_j(f,n)$ or $\Mat{W}(f)$, implying that there is a permutation ambiguity in the separated components for each frequency. Thus, the separated components of different frequency bins that originate from the same source need to be grouped together in order to complete source separation. This process is called permutation alignment \cite{Ikram2002a,Sawada2004a}.

\subsection{CVAE Source Model}
Incorporating an appropriate constraint into 
the power spectrogram $\Vec{V}_j = \{v_j(f,n)\}_{f,n}$
not only helps eliminate the permutation ambiguity but also provides a clue for estimating $\mathcal{W}$.
In the MVAE method, the complex spectrogram of a single source $\Vec{S}=\{s(f,n)\}_{f,n}$ is modeled using a CVAE \cite{Kingma2014semi}
conditioned on a class variable $\Mat{c}$.
Here, $\Mat{c}$ is a one-hot vector consisting of $C$ elements that indicates to which class the separated signal belongs. 
For example, speaker IDs can be used as the class category in multispeaker separation tasks, where the entries of $\Mat{c}$ will be $1$ at the index of a certain speaker and $0$ at all other indices.
\par 
Since the following applies to all sources, index $j$ will be omitted throughout this paragraph.
A CVAE consists of decoder and encoder networks.
The decoder network is designed to produce the parameters of the distribution $p_\theta^{*}(\Vec{S}|\Mat{z}, \Mat{c})$ of data $\Vec{S}$ given a latent space variable $\Mat{z}$ and a class variable $\Mat{c}$. 
The encoder network is designed to generate the parameters of
a conditional distribution $q_\phi^{*}(\Mat{z}|\Vec{S}, \Mat{c})$
that approximates the exact posterior $p_\theta^{*}(\Mat{z}|\Vec{S}, \Mat{c})$.
The goal of the CVAE training is to find the weight parameters in the encoder and decoder networks, namely $\theta$ and $\phi$, such that the encoder distribution $q_\phi^{*}(\Mat{z}|\Vec{S}, \Mat{c})$ becomes consistent with the posterior $p_\theta^{*}(\Mat{z}|\Vec{S}, \Mat{c}) \propto p_\theta^{*}(\Vec{S}|\Mat{z}, \Mat{c})p(\Mat{z})$. 
Note that the KL divergence between 
$q_\phi^{*}(\Mat{z}|\Vec{S}, \Mat{c})$ and $p_\theta^{*}(\Mat{z}|\Vec{S}, \Mat{c})$
is shown to be equal to 
the difference between 
the log marginal likelihood
$p_\theta^{*}(\Vec{S}|\Mat{c})=\int_{\Mat{z}} p(\Vec{S}|\Mat{z},\Mat{c})p(\Mat{z}) \,d \Mat{z}$
and its variational lower bound.
Hence, 
minimizing the KL divergence between  
$q_\phi^{*}(\Mat{z}|\Vec{S}, \Mat{c})$ and $p_\theta^{*}(\Mat{z}|\Vec{S}, \Mat{c})$
amounts to maximizing the following variational lower bound \cite{Kingma2014auto}:
\begin{align}
	\mathcal{J} &=  \E_{(\Vec{S}, \Mat{c})} \big[
	\E_{\Mat{z}\sim q_\phi^{*}(\Mat{z} |\Vec{S}, \Mat{c})} [\log p_\theta^{*}(\Vec{S}|\Mat{z}, \Mat{c}) ] \nonumber \\
	&~~~~~~~~~~~~~~~~~~~~~~~~~~
	-{\rm{KL}}[q_\phi^{*}(\Mat{z}|\Vec{S}, \Mat{c}) || p(\Mat{z})]\big],
\label{eq:cvae_loss}
\end{align}
where we have used $\E_{(\Vec{S}, \Mat{c})} [\cdot]$ to denote the sample mean of its argument over the training examples $\{\Vec{S}_m, \Mat{c}_m\}_{m=1}^M$, and ${\rm{KL}}[\cdot||\cdot]$ to denote the KL divergence. 
Although it is difficult to obtain an analytical form of the expectation $\E_{\Mat{z}\sim q_\phi^{*}(\Mat{z} |\Vec{S}, \Mat{c})}[\cdot]$ in the first term of $\mathcal{J}$, we can use a reparameterization trick \cite{Kingma2014auto} to obtain a form that allows us to compute the gradient with respect to $\phi$ using a Monte Carlo approximation.
Now, 
$q_\phi^{*}(\Mat{z}|\Vec{S}, \Mat{c})$, $p_\theta^{*}(\Vec{S}|\Mat{z}, \Mat{c})$, and $p(\Mat{z})$ are distributions that need to be modeled.
In the MVAE method, 
$p(\Mat{z})$ and $q_\phi^{*}(\Mat{z}|\Vec{S}, \Mat{c})$ are 
described as Gaussian distributions 
as with a regular CVAE:
\begin{align}
	p(\Mat{z}) &= \Normal(\Mat{z}|\Mat{0}, \Mat{I}), \\
	q_\phi^{*}(\Mat{z}|\Vec{S}, \Mat{c}) &= \Normal (\Mat{z}|\Mat{\mu}_\phi^{*} (\Vec{S}, \Mat{c}), \Diag (\Mat{\sigma}_\phi^{*}{}^2(\Vec{S}, \Mat{c}))), 
\label{eq:encoder1}
\end{align}
where $\Mat{\mu}_\phi^{*}(\Vec{S}, \Mat{c})$ and $\Mat{\sigma}_\phi^{*}{}^2(\Vec{S}, \Mat{c})$ are the encoder network outputs.
For stable training, the total energy of each training utterance is normalized to 1.
However, the energy of each source in a test mixture does not necessarily equal 1. 
To fill this gap, a scale factor $g$ is additionally introduced into the decoder distribution as a free parameter to be estimated at test time. 
Specifically, we use an expression of the decoder distribution with variance scaled by $g$.
Hence, the decoder distribution for 
the complex spectrogram $\Vec{S}$
is expressed as
\begin{align}
p_\theta^{*}(\Vec{S}|\Mat{z},\Mat{c},g)
= \prod_{f,n}
\Normal_{\C}(s(f, n)|0, g \sigma_\theta^{*}{}^2(f, n; \Mat{z}, \Mat{c})),
\label{eq:decoder1}
\end{align}
where $\sigma^{*}_\theta{}^2(f,n; \Mat{z}, \Mat{c})$ denotes the $(f, n)$th element of the decoder network output. 
\refeq{decoder1} is called the CVAE source model. 
\par
If we use the above CVAE source model to represent the complex spectrogram of the $j$th signal in a mixture signal, $\Mat{z}_j$, $\Mat{c}_j$, and $g_j$ are the unknown parameters to be estimated.
Since the CVAE source model is given in the same form as the LGM in \refeq{LGM} if we denote $g_j \sigma_\theta^{*}{}^2(f, n; \Mat{z}_j, \Mat{c}_j)$ by $v_j(f, n)$, 
using this as the generative model for each source gives the log-likelihood in the same form as \refeq{ll}. 

\subsection{Optimization Algorithm}
The goal of the source separation algorithm in the MVAE method is to maximize the 
posterior $p(\mathcal{W}, \Psi, \mathcal{G}|\mathcal{X})\propto p(\mathcal{X}|\mathcal{W}, \Psi, \mathcal{G})p(\Mat{z})p(\Mat{c})$
with respect to $\mathcal{W}$, $\Psi=\{\Mat{z}_j, \Mat{c}_j\}_j$, and $\mathcal{G}=\{g_j\}_j$, 
where $\Mat{z}$ is assumed to follow $\Normal(\Mat{0}, \Mat{I})$,
and $p(\Mat{c})$ is the empirical distribution of the training examples $\{\Mat{c}_m\}_m$, expressed as a multinomial distribution. 
Hence, the objective function is 
$\log p(\mathcal{X}|\mathcal{W}, \Psi, \mathcal{G}) + \log p(\Mat{z}) + \log p(\Mat{c})$.
A stationary point of this function 
can be found by iteratively updating $\mathcal{W}$, $\Psi$, and $\mathcal{G}$ so that 
the function value is guaranteed to be non-decreasing.
To update $\mathcal{W}$, we can use the IP method \cite{Ono2011stable}:
\begin{align}
\Mat{w}_j(f) &\leftarrow (\Mat{W}^\H(f)\Mat{\Sigma}_j(f))^{-1}\Mat{e}_j, \label{eq:ip1}\\
\Mat{w}_j(f) &\leftarrow \frac{\Mat{w}_j(f)}{\sqrt{\Mat{w}^\H_j(f)\Mat{\Sigma}_j(f)\Mat{w}_j(f)}},
\label{eq:ip2}
\end{align}
where $\Mat{\Sigma}_j(f)=\frac{1}{N}\sum_n \Mat{x}(f,n)\Mat{x}^\H(f,n)/v_j(f,n)$ and
$\Mat{e}_j$ denotes the $j$th column of an $I\times I$ identity matrix.
As for $\mathcal{G}$, the update rule
\begin{align}
	g_j \leftarrow \frac{1}{FN} \sum_{f,n} \frac{|\Mat{w}_j^\H (f)\Mat{x}(f,n)|^2}{\sigma_\theta^{*}{}^2(f,n;\Mat{z}_j, \Mat{c}_j)}
	\label{eq:scale}
\end{align} 
maximizes the objective function
with respect to $g_j$ when $\mathcal{W}$ and $\Psi$ are fixed.  
Under fixed $\mathcal{W}$ and $\mathcal{G}$,
the optimal $\Mat{z}_j$ and $\Mat{c}_j$ that maximize the objective function can be found using the gradient descent method. 
Note that $\Mat{c}_j$ can be updated under the sum-to-one constraint by inserting an appropriately designed softmax layer that outputs $\Mat{c}_j$.
\par
One important feature of VAE in general is its generalization capability, namely the ability to learn the distribution of unseen data.
Thanks to this feature, we expect that the CVAE source model trained on speech samples of sufficiently many speakers can generalize somewhat well to the spectrograms of unknown speakers, thus allowing the above algorithm to handle speaker-independent scenarios reasonably well.
Another advantage is that it is guaranteed to converge to a stationary point, making it easy to handle in practical use. 
However, the downside is that the backpropagation algorithm required for each iteration can be computationally expensive.

\section{FastMVAE}
\label{sec:FastMVAE}

\subsection{ACVAE Source Model}

\begin{figure*}[t]
\centering
\includegraphics[width=\linewidth]{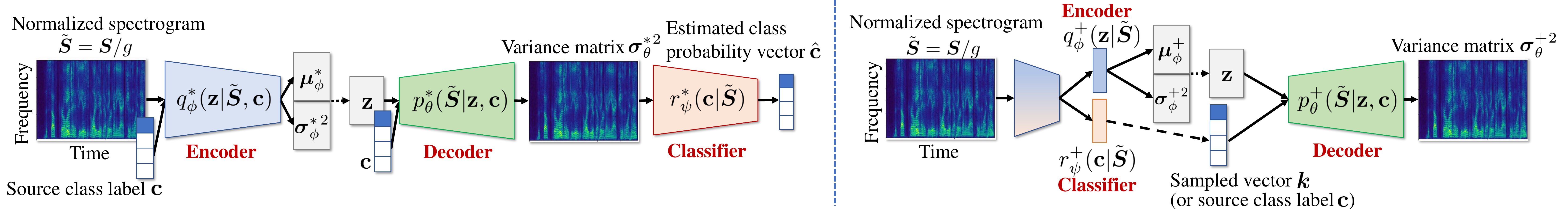} 
\caption{Illustration of the ACVAE model in FastMVAE (left) and the ChimeraACVAE model in FastMVAE2 (right). We use $\tilde{\Vec{S}}=\Vec{S}/g$ to denote a normalized spectrogram and omit $g=1$ in encoder, decoder, and classifier distributions.}
\label{fig:acvae}
\end{figure*}

The motivation behind the FastMVAE method is to accelerate the process of updating $\Psi$. 
Under fixed $\mathcal{W}$ and $\mathcal{G}$, the objective function of the MVAE method is equal to the sum of 
$\log p(\Mat{z}_j, \Mat{c}_j | \Vec{S}_j, g_j)$ 
up to a constant, 
where $\Vec{S}_j$ is the set $\{ \Mat{w}_j^\H (f)\Mat{x}(f,n) \}_{f,n}$, namely 
the complex spectrogram of the signal separated from the observed signal using the current estimate of $\mathcal{W}$.
The idea of the FastMVAE method is to express
this posterior as $p(\Mat{z}_j, \Mat{c}_j|\Vec{S}_j, g_j)=p(\Mat{z}_j|\Vec{S}_j, \Mat{c}_j, g_j)p(\Mat{c}_j|\Vec{S}_j, g_j)$ and use two trainable networks to approximate these two conditional distributions.
Once these networks have been trained, an approximation of the maximum point of the posterior $p(\Mat{z}_j, \Mat{c}_j | \Vec{S}_j, g_j)$ can be obtained by finding the maximum points of the two approximate distributions.
\par
To obtain approximations of the two conditional distributions, the FastMVAE method employs the idea of ACVAE training \cite{Kameoka2019acvae}. 
ACVAE is a CVAE variant that incorporates the expectation of
the mutual information \cite{Chen2016infogan} 
\begin{align}
&I (\Mat{c}, \Vec{S}|\Mat{z})  \nonumber\\
&=\E_{\Mat{c} \sim p_D(\Mat{c}), \Vec{S}\sim p_\theta(\Vec{S}|\Mat{z}, \Mat{c}), \Mat{c}'\sim p(\Mat{c}|\Vec{S})}
[\log p(\Mat{c}'|\Vec{S})]
+H(\Mat{c}),
\end{align}
into the training criterion with the aim of making the decoder output as correlated as possible with the class variable $\Mat{c}$.
Here, $p_D(\Mat{c})$ is the empirical discrete distribution of the samples of $\Mat{c}$ in the training set and $H(\Mat{c})$ represents the entropy of $\Mat{c}$, which can be regarded as a constant.
Since it is difficult to express $I (\Mat{c}, \Vec{S}|\Mat{z})$ in analytical form, rather than using it directly, ACVAE uses its variational lower bound 
\begin{align}
\label{eq:auxiliary} 
	&\mathcal{L} = \E_{(\Vec{S}, \Mat{c}'), \Mat{z} \sim q_\phi^{*}(\Mat{z}|\Vec{S}, \Mat{c}')
	[
	\E_{\Mat{c}, \Vec{S}\sim p_\theta^{*}(\Vec{S}|\Mat{z}, \Mat{c})} 
	[\log r_\psi^{*}(\Mat{c}|\Vec{S}, g)]]}
\end{align}
defined using a variational distribution $r_\psi^{*}(\Mat{c}|\Vec{S}, g)=\rm{Mult}(\Mat{c}|\Mat{\rho}_\psi^{*}(\Vec{S}/g))$ for optimization,
where $\E_{(\Vec{S},\Mat{c}')}[\cdot]$ is equivalent to $\E_{(\Vec{S}, \Mat{c})}[\cdot]$, $\E_{\Mat{c}}[\cdot]$ denotes the mean of its argument over all one-hot vectors $\Mat{c}\sim p_D(\Mat{c})$, which can be approximated by a Monte Carlo approximation, and $\E_{\Mat{z} \sim q_\phi^{*}(\Mat{z}|\Vec{S}, \Mat{c})}[\cdot]$ and $\E_{\Vec{S}\sim p_\theta^{*}(\Vec{S}|\Mat{z}, \Mat{c}')}[\cdot]$ are approached by a Monte Carlo approximation after reparameterization tricks.
Here, 
$\rm{Mult}(\Mat{c}|\Mat{\rho})\propto \prod_i \rho_i^{c_i}$ denotes a multinomial distribution, where $\Mat{c}=[c_1,\ldots,c_I]^{\mathsf T}$ and $\Mat{\rho}=[\rho_1,\ldots,\rho_I]^{\mathsf T}$. 
$\Mat{\rho}_\psi^{*}(\Vec{S}/g)$ is a neural network 
that takes $\Vec{S}$ normalized by $g$ as an input and produces a probability vector consisting of $C$ elements that sum to 1.
$r_\psi^{*}(\Mat{c}|\Vec{S}, g)$ is an auxiliary classifier.
Since the exact bound is obtained when $r_\psi^{*}(\Mat{c}|\Vec{S}, g) = p(\Mat{c}|\Vec{S}, g)$, the trained auxiliary classifier $r_\psi^{*}(\Mat{c}|\Vec{S}, g)$ is expected to be a good approximation of the distribution $p(\Mat{c}|\Vec{S}, g)$ of interest.
ACVAE also uses the negative cross-entropy
\begin{align}
	\mathcal{I}=\E_{(\Vec{S}, \Vec{c})}[\log r_\psi^{*}(\Vec{c}|\Vec{S}, g)]
\end{align} 
as the training criterion. 
Therefore, the entire training criterion to be maximized is given by 
\begin{align}
	\mathcal{J} + \lambda_\mathcal{L}\mathcal{L}
	+ \lambda_\mathcal{I}\mathcal{I},
	\label{eq:acvae}
\end{align}
where $\lambda_\mathcal{L}, \lambda_\mathcal{I}\geq 0$ denote the regularization weights that weight the importance of the regularization terms.  
The set of the networks trained in this way using the spectrograms of the training utterances is called the {\it ACVAE source model}.
An illustration of ACVAE is shown on the left of \reffig{acvae}.

\subsection{Optimization Algorithm}
\begin{algorithm}[t]
\caption{FastMVAE algorithm w/ PoE}
\begin{algorithmic}[1]
\REQUIRE {Network parameter $\theta$, $\phi$, $\psi$ trained using \refeq{acvae}, observed mixture signal $\Mat{x}(f, n)$, iteration number $\mathscr{L}$, weight parameter $\alpha$}
\STATE $\text{randomly initialize }\mathcal{W}\text{, }\Psi$
\FOR {$\ell = 1$ to $\mathscr{L}$}
	\FOR {$j=1$ to $J$}
		\STATE $y_j(f, n) = \Mat{w}_j^\H(f) \Mat{x}(f,n)$
		\STATE $\text{(updating source model parameters)}$
		\STATE $\text{initialize }g_j\text{ using }\refeq{scale}$
		\STATE $\text{normalize } \Vec{\bar{S}}_j = \{y_j(f, n) / g_j\}_{f, n}$
		\STATE $\text{update }\Mat{c}_j\text{ using }\refeq{update_c_cont}$
		\STATE $\text{update }\Mat{z}_j\text{ using }\refeq{update_z}$
		\STATE $\text{compute } \sigma_j^{*}{}^2(f, n; \Mat{z}_j, \Mat{c}_j, g_j=1, \theta)$
		\STATE $\text{update }g_j\text{ using }\refeq{scale}$
		\STATE $\text{compute }v_j(f, n) = g_j \cdot  \sigma_j^{*}{}^2(f, n; \Mat{z}_j, \Mat{c}_j, g_j=1, \theta)$
		\STATE $\text{(updating separation matrices)}$
		\FOR {$f=1$ to $F$}
    		\STATE $\text{update }\Mat{w}_j(f)\text{ by IP method with }\refeq{ip1}, \refeq{ip2}$
		\ENDFOR
	\ENDFOR
\ENDFOR
\end{algorithmic}
\end{algorithm}

After ACVAE training, we achieve $p(\Mat{z}_j, \Mat{c}_j|\Vec{S}_j, g_j) \approx r_\psi^{*}(\Mat{c}_j|\Vec{S}_j, g_j) q_\phi^{*}(\Mat{z}_j|\Vec{S}_j, \Mat{c}_j, g_j)$.
Since the maximum points of 
$r_\psi^{*}(\Mat{c}_j|\Vec{S}_j, g_j)$ and $q_\phi^{*}(\Mat{z}_j|\Vec{S}_j, \Mat{c}_j, g_j)$
can be found through the forward passes of the auxiliary classifier and encoder, respectively, we can quickly find an approximate solution to 
$(\Mat{z}_j,\Mat{c}_j) = \Argmax_{\Mat{z}_j,\Mat{c}_j} p(\Mat{z}_j, \Mat{c}_j|\Vec{S}_j, g_j)$ without resorting to gradient descent updates.
Specifically, $\Mat{c}_j$ is given as the probability vector produced by the auxiliary classifier network:
\begin{align} 
 	\Mat{c}_j \leftarrow \Mat{\rho}_{\psi}^{*}(\Vec{S}_j/g_j),
 	\label{eq:update_c_cont}
\end{align}
and 
$\Mat{z}_j$ is given as the mean of the encoder distribution: 
\begin{align}
\Mat{z}_j \leftarrow \Mat{\mu}_\phi^{*}(\Vec{S}_j/g_j, \Mat{c}_j).
\end{align}
Here, if the $j$th separated signal corresponds to a speaker unseen in the training set, the elements of \refeq{update_c_cont} can be interpreted as  quantities indicating how similar that speaker is to all the speakers in the training set. If the signal of any speaker can be assumed to be expressed as a point in the manifold spanned by all the speakers in the training set, our algorithm is expected to be able to handle even mixtures of unknown speakers.
\par
However, our preliminary experiments revealed that directly using the mean of the encoder distribution tends to degrade source separation performance for speakers not included in the training set. 
To stabilize the inference for unknown speakers, we previously proposed reapplying the prior $p(\Mat{z}_j)$ to the encoder output based on the PoE framework \cite{Hinton2002training}to ensure that $\Mat{z}_j$ will not be updated to an outlier. 
Namely, the prior $p(\Mat{z}_j)$ is redefined as the product of two distributions with respect to $\Mat{z}_j$, namely,
$\Argmax_{\Mat{z}_j} p(\Mat{z}_j|\Vec{S}_j, \Mat{c}_j, g_j)p(\Mat{z}_j)^\alpha$.
Accordingly, the modified update rule of $\Mat{z}_j$ is given as
\begin{align}
\Mat{z}_j \leftarrow \Mat{\Sigma}^{-1}_{\phi,j}(\Mat{\Sigma}^{-1}_{\phi, j} +\alpha\Mat{I})^{-1}\Mat{\mu}_\phi^{*}(\Vec{S}_j/g_j, \Mat{c}_j).
\label{eq:update_z} 
\end{align}
Here, $\alpha$ is a parameter that weights the importance of the prior $p(\Mat{z}_j)$ in the inference, and $\Mat{\Sigma}_{\phi,j}=\Diag(\Mat{\sigma}_\phi^{*}{}^2(\Vec{S}_j/g_j, \Mat{c}_j))$. 
Note that \refeq{update_z} reduces to the mean of the encoder distribution when $\alpha=0$.
The algorithm of the FastMVAE method is summarized in {\bf Algorithm 1}.

\section{Proposed: FastMVAE2}
\label{sec:proposed}
While the FastMVAE method can significantly reduce the computation time compared to the MVAE method, its source separation accuracy has been confirmed to be somewhat less than that of the MVAE method \cite{Li2020fast}. We believe that this is due to the limitations of the generalization capabilities of the encoder and classifier obtained from the ACVAE training. In this paper, we propose introducing a new model architecture and training scheme to overcome these limitations, rather than implementing a heuristic solution at the inference stage. 

\subsection{ChimeraACVAE source model}
We first describe our motivation and ideas for developing an improved version of the ACVAE source model, which we call the {\it ``ChimeraACVAE'' source model}.

\subsubsection{Multitask encoder}
When performing source separation, it is desirable that the speaker identity of each separated signal does not change over time. This is because a change of the identity of each separated signal means a failure in source separation. However, constraining the identity not to change is not an easy task if the decoder is not conditioned on $\Mat{c}$ (as in a regular VAE), since it will be trained so that $\Mat{z}$ becomes an entangled mixture of linguistic and speaker-identity information. 
	In contrast, conditioning the decoder on $\Mat{c}$ is expected to promote disentanglement between $\Mat{z}$ and $\Mat{c}$ so that $\Mat{z}$ represents only the linguistic information and $\Mat{c}$ represents only the speaker identity.
	This allows our source separation system to always ensure that the speaker identity of each separated signal is time-invariant. Thus, it is essential for the decoder to remain conditioned on $\Mat{c}$, and it is the encoder that we propose to modify. Specifically, we unify the encoder and auxiliary classifier into a single network with two branches that output the parameters of the encoder distribution $q_\phi^{+}(\Mat{z}|\Vec{S}, g)=\mathcal{N}(\Mat{z}|\Mat{\mu}_{\phi}^{+}(\Vec{S}/g), {\rm diag}(\Mat{\sigma}_{\phi}^{+}{}^2(\Vec{S}/g)))$ and those of the class distribution $r_\psi^{+}(\Mat{c}|\Vec{S}, g) = {\rm Mult}(\Mat{c}|\Mat{\rho}_{\psi}^{+}(\Vec{S}/g))$, respectively. 
Here, the latent variable $\Mat{z}$ and speaker identity $\Mat{c}$ are assumed to be conditionally independent. 
We believe that the main reason for the performance degradation in FastMVAE under the speaker-independent condition is the cascade structure of the classifier and encoder, where errors in the classifier directly affect the outputs of the encoder.
The conditional independence assumption in the ChimeraACVAE source model allows us to parallelize the processes by the classifier and encoder and prevent error propagation. 
Furthermore, the sharing of the layers in the unified encoder network is expected to improve the generalization capability through multitask learning.
	
\subsubsection{Network details}
The original ACVAE source model is designed to include batch normalization layers in its networks.
However, since the computation of batch normalization depends on the mini-batch size, the learned parameters may be suboptimal in inference situations where the number of sources differs from the mini-batch size during training. 
To avoid inconsistencies in computation during training and inference, we replace batch normalization \cite{batchnormalization} with layer normalization \cite{Ba2016layer}.
In addition, we use a sigmoid linear unit (SiLU) \cite{SiLU} instead of a gated linear unit (GLU) \cite{Dauphin2017language} to reduce model size. SiLU, also known as the swish activation function, is a self-gated activation function, which can be expressed as
	\begin{align}
	\label{eq:silu}
	\mathbbm{O}_l = (\mathbbm{O}_{l-1} \ast \mathbbm{W}_l+\mathbbm{b}_l) \otimes \sigma (\mathbbm{O}_{l-1} \ast \mathbbm{W}_l+\mathbbm{b}_l)
	\end{align} 
when applied to a convolution layer.
Here, $\mathbbm{W}_l$ and $\mathbbm{b}_l$ are weight and bias parameters of the $l$th layer, and $\mathbbm{O}_l$ and $\mathbbm{O}_{l-1}$ denote the output and input of the $l$th layer, respectively. $\otimes$ denotes element-wise multiplication, and $\sigma(\cdot)$ is the sigmoid function. 
Both SiLU and GLU are data-driven gates, which control the information passed in the hierarchy. Unlike GLU, where the linear and gate functions are parametrized separately, SiLU uses the same parameters to represent them. This halves the number of parameters in a single layer.
\par
\begin{figure}[t!]
\centering
\includegraphics[width=\linewidth]{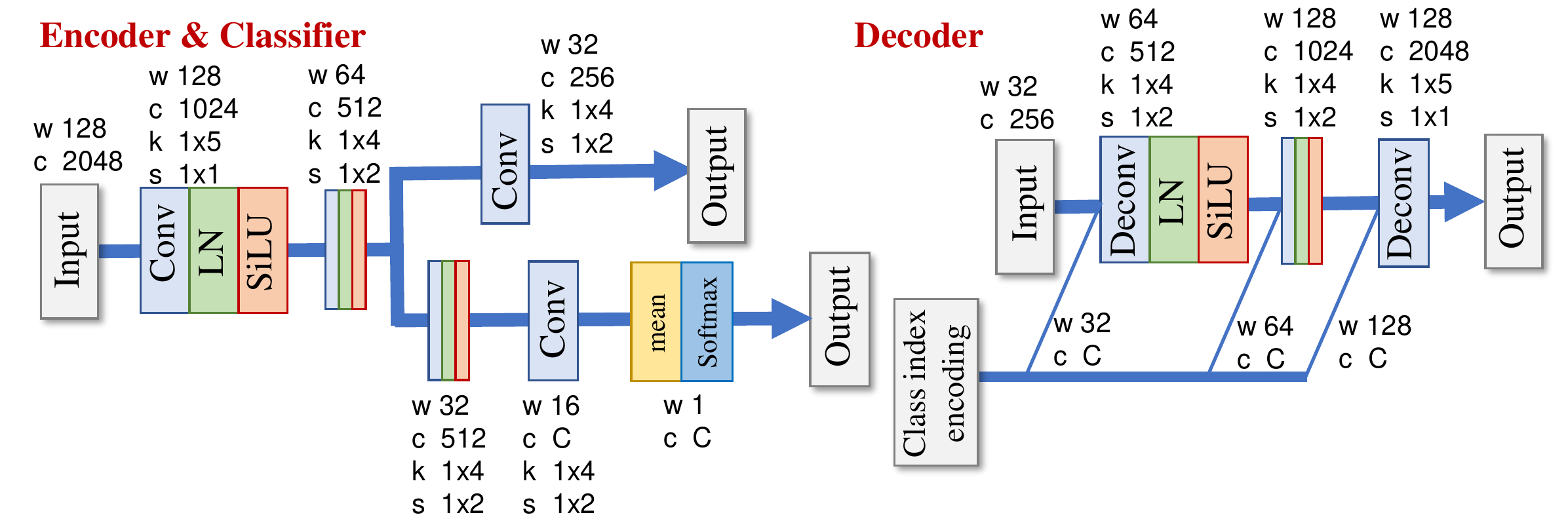} 
\caption{
Network architectures of the unified encoder and decoder in the ChimeraACVAE source model. The inputs and outputs are assumed to be vector sequences. A spectrogram is interpreted as a sequence of spectra, with frequency regarded as the channel dimension. 
``w''denotes the length of the input sequence. 
``Conv'' and ``Deconv'' denote one-dimensional convolution and deconvolution, respectively, where ``c'', ``k'', and ``s'' denote the channel number, kernel size, and stride size, respectively.
``LN'' and ``SiLU'' stand for the layer normalization and sigmoid linear unit, respectively.
``mean'' denotes the operation of averaging the input sequence along the time direction, and ``softmax'' denotes the operation of applying a softmax function to the input vector.
In the decoder, the ``class index encoding'' $\Vec{c}$ is concatenated to the input of each deconvolution layer along the channel direction after being repeated along the time direction so that it has the shape compatible with the input.
}
\label{fig:nn}
\end{figure}

An illustration of the proposed ChimeraACVAE source model is shown on the right in \reffig{acvae}, and the network architecture used to configure the model is shown in \reffig{nn}. 
\reftab{num} shows the number of the parameters of the CVAE, ACVAE, and ChimeraACVAE models used in the following experiments. 
Note that the number of parameters depend on the number of speakers in the training dataset.
As can be seen from this comparison, the ChimeraACVAE source model with the above modifications has reduced the number of parameters to about 40\% of the original ACVAE source model, which is even smaller than that in the CVAE model used in the MVAE method.

\begin{table}[t!]
\centering
\caption{Number of parameters of CVAE, ACVAE, and ChimeraACVAE model used in the experiments.}
\begin{tabular}{lrr}
\hline \hline
\multirow{2}{*}{Model}    & \multicolumn{2}{c}{Number of parameters [M]}  \\
& Spk-dep & Spk-ind \\
\hline\hline
CVAE      &     10.6 & 12.5  \\
ACVAE  &                17.0 & 18.9\\
ChimeraACVAE &     7.0  & 7.9\\
\hline   
\end{tabular}
\label{tab:num}
\end{table}

\subsection{Training criterion based on KD}
\label{subsec:criterion}

\begin{figure}[t]
\centering
\includegraphics[width=\linewidth]{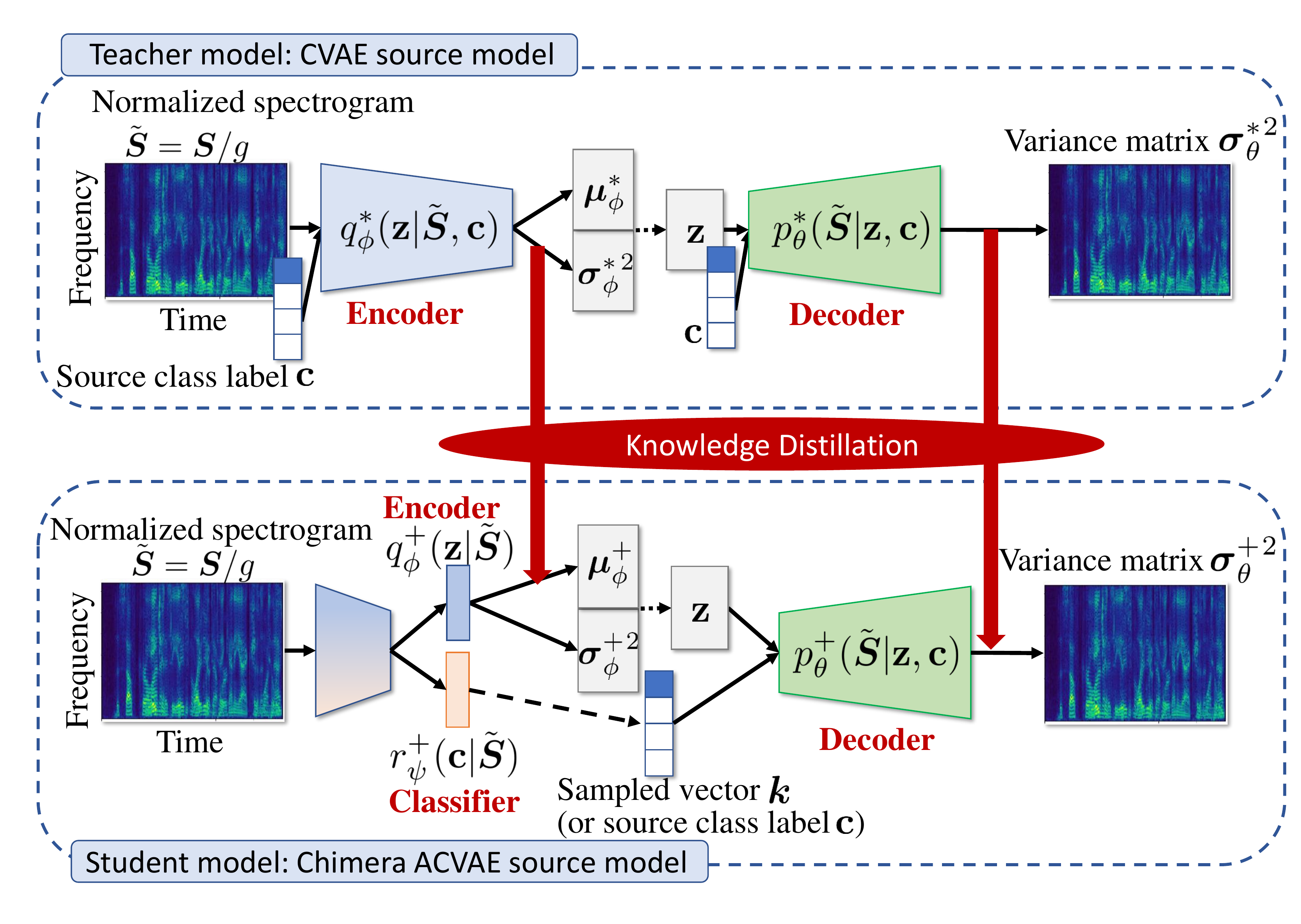} 
\caption{Illustration of the response-based KD from a pre-trained CVAE source model to the ChimeraACVAE source model. We use $\tilde{\Vec{S}}=\Vec{S}/g$ to denote a normalized spectrogram. 
Note that $g=1$ is omitted from the expressions of the encoder, decoder, decoder distributions owing to space limitations.
}
\label{fig:kd}
\end{figure}

Since the latent variable $\Mat{z}$ no longer depends on $\Mat{c}$, we must first rewrite the training loss of ACVAE, i.e., \refeq{acvae}, by replacing $q_\phi^{*}(\Mat{z}|\Vec{S}, \Mat{c})$ with $q_\phi^{+}(\Mat{z}|\Vec{S})$. Note that we omit $g$ in this subsection, assuming that $g$ is set to 1 and normalized spectrograms are used during training.
Thus, the reformulated training criteria are given as
\begin{align}
	&\mathcal{J} =  \E_{\Vec{S}, \Mat{c}}\big[
	\E_{\Mat{z}\sim q_\phi^{+}(\Mat{z} |\Vec{S})} [\log p_\theta^{+}(\Vec{S}|\Mat{z}, \Mat{c}) ] 
	-{\rm{KL}}[q_\phi^{+}(\Mat{z}|\Vec{S}) || p(\Mat{z})]\big], 
	\label{eq:modified_acvae_1}\\
	&\mathcal{L} = \E_{\Vec{S}', \Mat{z}\sim q_\phi^{+}(\Mat{z}|\Vec{S}')}
	\big[
	\E_{\Mat{c}, \Vec{S} \sim p_\theta^{+}(\Vec{S} |\Mat{z}, \Mat{c})} 
	[\log r_\psi^{+}(\Mat{c}|\Vec{S})]\big],
	\label{eq:modified_acvae_2}\\
	&\mathcal{I}=\E_{\Vec{S}, \Vec{c}}[\log r_\psi^{+}(\Vec{c}|\Vec{S})].
	\label{eq:modified_acvae_3}
\end{align}
Here, $\E_{\Mat{S}'}[\cdot]$ in \refeq{modified_acvae_2} denote the mean of the arguments over all spectrograms $\Vec{S}' \sim p_D(\Vec{S})$ in the training dataset. The superscript $^{+}$ is used to distinguish the networks in the ChimeraACVAE model from those in the original ACVAE model superscripted with $^{*}$. 
\par
Unlike in the training phase, where the class label $\Mat{c}$ is known and given, in the separation phase, the spectrogram $\Vec{S}$ needs to be constructed using the estimated $\Mat{z}$ and $\Mat{c}$. 
Therefore, it is reasonable to simulate this situation in the training phase as well.
Namely, we consider not only the reconstruction error defined using the given label $\Mat{c}$ but also the reconstruction error defined using the estimated $\Mat{c}\sim r_\psi^{+}(\Mat{c}|\Vec{S})$.
Thus, we propose including
\begin{align}
&\mathcal{J}' = \E_{\Vec{S}, \Mat{z}\sim q_\phi^{+}(\Mat{z} |\Vec{S}), \Mat{c}\sim r_\psi^{+}(\Mat{c}|\Vec{s})} [\log p_\theta^{+}(\Vec{S}|\Mat{z}, \Mat{c})], 
\label{eq:est_l_1} \\
&\mathcal{L}' = \E_{\Vec{S}', \Mat{z}\sim q_\phi^{+}(\Mat{z}|\Vec{S}'), \Mat{c}\sim r_\psi^{+}(\Mat{c}|\Vec{S}')}
[\E_{\Vec{S}\sim p_\theta^{+}(\Vec{S}|\Mat{z}, \Mat{c})} 
[\log r_\psi^{+}(\Mat{c}|\Vec{S})]],
\label{eq:est_l_2}
\end{align}
in the training objective. Here, it should be noted that both $\mathcal{J}'$ and $\mathcal{L}'$ involve expectations over $\Mat{c}\sim r_\psi^{+}(\Mat{c}|\Vec{S}')$. 
However, there is currently no known reparametrization trick that can be applied to random variables that follow multinomial distributions. 
Instead, we use the Gumbel-Softmax (GS) distribution as an approximation to the  multinomial distribution, which allows the use of the reparameterization trick \cite{Jang2017categorical,Maddison2017the}.
The GS distribution of a continuous multivariate variable $\Vec{k}=[k_1, \ldots, k_{\rm I}]^\T$ is defined as
\begin{align}
p_{\Vec{\rho}, \tau}(\Vec{k})=\Gamma({\rm I})\tau^{{\rm I}-1}
\Bigg(\sum_{{\rm i}=1}^{\rm I} \rho_{\rm i} / k_{\rm i}^{\tau}\Bigg)^{-{\rm I}}
\prod_{{\rm i}=1}^{\rm I} \big(\rho_{\rm i}/k_{\rm i}^{\tau+1}\big).
\label{eq:gumbel-softmax}
\end{align}
This expression is derived analytically as a distribution that is followed by the variables 
\begin{align}
    k_{\rm i} = \frac{\exp ((\log \rho_{\rm i} + {\rm g}_{\rm i})/\tau)}
    {\sum_{{\rm i}'=1}^{\rm I} \exp ((\log \rho_{\rm i} + {\rm g}_{\rm i})/\tau)}~~
    (i=1,\ldots,I)
\end{align}
where ${\rm g}_{\rm i},~{\rm i}=1,\ldots, {\rm I}$ are Gumbel samples drawn independently and identically from ${\rm Gumbel}(0, 1)$,
$\Vec{\rho}$ is the class probability vector produced by the classifier, 
and $\tau$ is called the softmax temperature. 
Here, it is important to note that \refeq{gumbel-softmax} is shown to become identical to $r_\psi^{+}(\Vec{k}|\Vec{S}')$ as $\tau$ approaches 0.
By replacing $r_\psi^{+}(\Vec{k}|\Vec{S}')$ with \refeq{gumbel-softmax}, \refeqs{est_l_1}{est_l_2} can be approximated as 
\begin{align}
\label{eq:est_l_5} 
    \mathcal{J}'_{{\rm GS}}&=  
	\E_{\Vec{S}, \Mat{z}\sim q_\phi^{+}(\Mat{z} |\Vec{S}), \Vec{k}\sim p_{\hat{\Vec{\rho}}, \tau}(\Vec{k})} 
	\big[ \log p_\theta^{+}(\Vec{S}|\Mat{z}, \Vec{k}) \big],
	\\
	\mathcal{L}'_{\rm GS}&= 
	\E_{\Vec{S}', \Mat{z}\sim q_\phi^{+}(\Mat{z}|\Vec{S}'),\Vec{k}\sim p_{\hat{\Vec{\rho}}, \tau}(\Vec{k}),
	\Vec{S}\sim p_\theta^{+}(\Vec{S}|\Mat{z}, \Vec{k})} 
	\big[\log r_\psi^{+}(\Vec{k}|\Vec{S}) \big].
	\label{eq:est_l_6}
\end{align}
Unlike the original expressions, these expressions allow the computations of the derivatives with respect to $\psi$ using the reparameterization trick.
\par
With the reduced number of model parameters, the challenge is how to make the ChimeraACVAE model have a high generalization capability. 
To this end, we further introduce training criteria derived based on the KD \cite{distillation} using a pretrained CVAE model as the teacher model.
KD, also known as teacher-student learning, is a technique to transfer the knowledge from a teacher model to a student model, originally proposed for model compression \cite{distillation} and later shown to improve the generalization capability of the student model \cite{Shen2018feature}. 
There are three types of knowledge that can be transferred between models: response-based knowledge, feature-based knowledge, and relation-based knowledge. These refer to the knowledge of the last output layer, the knowledge of each output layer, and the knowledge of the relationship between layers, respectively. 
Since 
the networks in both the teacher and student models are reasonably shallow,
we consider response-based KD to be sufficient, as it requires a minimal increase in training cost.
\par
Specifically, we transfer the knowledge of the distributions of the latent variable $q_\phi^{*}(\Mat{z}|\Vec{S}, \Mat{c})$ and the complex spectrograms $p_\theta^{*}(\Vec{S}|\Mat{z}, \Mat{c})$ learned by the CVAE model into the ChimeraACVAE model by using these distributions as priors. We use the KL divergences to measure the differences between the distributions estimated by a student model and the pretrained teacher model such that
\begin{align}
&\mathcal{K}_{\Mat{z}}= \E_{\Vec{S}, \Mat{c}}\big[{\rm{KL}}[q_\phi^{*}(\Mat{z}|\Vec{S}, \Mat{c}) || q_\phi^{+}(\Mat{z}|\Vec{S})]\big], 
\label{eq:KD_z} \\
&\mathcal{K}_{\Vec{S}}
= \E_{\Vec{S}, \Mat{c}, \Mat{z}^{*}\sim q_\phi^{*}(\Mat{z}|\Vec{S}, \Mat{c}), \Mat{z}^{+}\sim q_\phi^{+}(\Mat{z}|\Vec{S})}  \nonumber \\ 
&~~~~~~~~~~~~~~~~~~~~
\big[{\rm{KL}}[p_\theta^{*}(\Vec{S}|\Mat{z}^{*}, \Mat{c}) || p_\theta^{+}(\Vec{S}|\Mat{z}^{+}, \Mat{c}) ]\big], 
\label{eq:KD_true}
\\
&\mathcal{K}^{'}{}_{\Vec{S}
}
= \E_{\Vec{S},\Mat{c}, \Mat{z}^{*}\sim q_\phi^{*}(\Mat{z}|\Vec{S}, \Mat{c}), \Mat{z}^{+}\sim q_\phi^{+}(\Mat{z}|\Vec{S}), \Vec{k}\sim p_{\hat{\Vec{\rho}}, \tau}(\Vec{k})} \nonumber \\
&~~~~~~~~~~~~~~~~~~~~~
\big[{\rm{KL}}[p_\theta^{*}(\Vec{S}|\Mat{z}^{*}, \Mat{c}) || p_\theta^{+}(\Vec{S}|\Mat{z}^{+}, \Vec{k}) ]\big].
\label{eq:KD_est_GS}
\end{align}
Here, \refeq{KD_est_GS} is a criterion that measures the difference between the teacher distribution and decoder distribution computed using the GS distribution.
An illustration of KD for training the ChimeraACVAE model is shown in \reffig{kd}.
\par
The total training criterion of the ChimeraACVAE is a weighted linear combination of the above-mentioned criteria:
\begin{align}
&\mathcal{J} + \lambda_{\mathcal{L}}\mathcal{L} + \lambda_{\mathcal{I}}\mathcal{I}
+\lambda_{\mathcal{J}'}\mathcal{J}'_{\rm GS} + \lambda_{\mathcal{L}'}\mathcal{L}'_{\rm GS} \nonumber \\
&~~~~~~~~~~~~~~~~~~~~~
- \lambda_{\mathcal{K}_{\Mat{z}}}\mathcal{K}_{\Mat{z}}
-\lambda_{\mathcal{K}_{\Vec{S}}
}\mathcal{K}_{\Vec{S}}
-\lambda_{\mathcal{K}^{'}{}_{\Vec{S}
}
}\mathcal{K}^{'}{}_{\Vec{S}
}
.
\end{align}
Here, $\lambda_{\ast}$
denotes a non-negative parameter that weighs the importance of that term.
\par
With the trained ChimeraACVAE source model, we can use the same procedure as {\bf Algorithm 1} to perform source separation. We call it {\it FastMVAE2} 
to distinguish it from the method using the ACVAE source model. 
Note that in FastMVAE2, the PoE-based update rule is no longer required thanks to the improved generalization capability, but of course it can be used in addition.

\section{Experimental evaluations}
\label{sec:experiment}

To evaluate the effectiveness of the proposed training procedure, we compare the source separation performance in speaker-dependent and speaker-independent situations. 

\subsection{Datasets}
For the speaker-dependent source separation experiment, we used speech utterances of two male speakers (SM1, SM2) and two female speakers (SF1, SF2) excerpted from the Voice Conversion Challenge (VCC) 2018 dataset \cite{Lorenzo2018voice} . 
The audio files for each speaker were about seven minutes long and manually segmented into 116 short sentences, where 81 and 35 sentences (about five and two minutes long, respectively) served as training and test sets, respectively.
We used two-channel mixture signals of two sources as the test data, which were synthesized using simulated room impulse responses (RIRs) generated using the image method \cite{Allen1979image} and real RIRs measured in an anechoic room (ANE) and an echo room (E2A).  
The reverberation times ($RT_{60}$) \cite{Schroeder1965new} of the simulated RIRs were set at 78 and 351 ms, which were controlled by setting the reflection coefficient of the walls at 0.20 and 0.80, respectively.
For the measured RIRs, we used the data included in the RWCP Sound Scene Database in Real Acoustic Environments \cite{Nakamura1999sound}. 
The $RT_{60}$ of ANE and E2A were 173 and 225 ms, respectively.
The test data included four pairs of speakers, i.e., SF1+SF2, SF1+SM1, SM1+SM2, and SF2+SM2.  For each speaker pair, we generated ten mixture signals. 
Hence, there were a total of 40 test signals for each reverberation condition, each of which was about four to seven seconds long.
\par
The datasets for the speaker-independent experiment were generated in the same way by using the Wall Street Journal (WSJ0) corpus \cite{Garofolo1993}.  
All the utterances in WSJ0 folder $\tt si\_tr\_s$ (around 25 hours) were used as the training set, which consists of 101 speakers in total. 
A test set was created by randomly mixing two different speakers selected from the WSJ0 folders $\tt si\_dt\_05$ and $\tt si\_et\_05$, where the number of speakers was 18. We generated test data using simulated RIRs with $RT_{60}=78$ ms and $RT_{60}=351$ ms, where 100 mixture signals were generated under each reverberation condition. 
All the speech signals were resampled at 16 kHz. 
The STFT and inverse STFT were calculated by using a Hamming window with a length of 128 ms and half overlap. 

\subsection{Experimental settings}
We chose ILRMA \cite{Kitamura2016determined}, the MVAE method \cite{Kameoka2019supervised}\footnote{Code: https://github.com/lili-0805/MVAE}
, and the FastMVAE method \cite{Li2020fast} as the baseline methods for both the speaker-dependent and speaker-independent cases, and IDLMA \cite{Mogami2018independent} as another baseline method only for the speaker-dependent scenario.
For all the methods, the parameter optimization algorithms were run for 60 iterations, and the separation matrix $\Mat{W}(f)$ was initialized with an identity matrix.
\par
We set the basis number of ILRMA at 2, which is the optimal setting for speech separation. 
For IDLMA, we used the same network architecture and training settings as those in \cite{Mogami2018independent} except for the optimization algorithm, where we used Adam \cite{Kingma2015adam} instead of Adadelta \cite{Adadelta}. 
Note that unlike other methods where speaker information is estimated, IDLMA requires speaker information in order to properly select the corresponding pre-trained network.
The network architectures for the CVAE and ACVAE source models were the same as those used in \cite{Li2020fast}, where the encoder consisted of 2 convolutional layers using GLU following a regular convolutional layer, the decoder consisted of 2 deconvolutional layers using GLU following a regular deconvolutional layer, and the classifier consisted of 3 convolutional layers using GLU following a regular convolutional layer. All the GLU layers used batch normalization to stabilize the training. Adam was used to train the networks and estimate $\Mat{z}_j$ and $\Mat{c}_j$ in the MVAE method. 
In the training of ChimeraACVAE, the weight parameters were empirically set with the KD criterion $\mathcal{K}_{\Mat{z}}$ as 10 and the rest as 1. The temperature $\tau$ for the GS distribution was set at 1.
\par
We calculated the source-to-distortions ratio (SDR), source-to-interferences ratio (SIR), and sources-to-artifacts ratio (SAR) \cite{Vincent2006performance} to evaluate the source separation performance, and used perceptual evaluation of speech quality (PESQ)\footnote{Code: https://github.com/vBaiCai/python-pesq} \cite{Rix2001perceptual} and short-time objective intelligibility (STOI) \footnote{Code: https://github.com/mpariente/pystoi} \cite{Taal2010a} to ascertain the speech quality and intelligibility of the separated waveforms.

\subsection{Multi-speaker separation performance}
\begin{table}[t!]
\setlength{\tabcolsep}{10pt}
\centering
\caption{Evaluated models and corresponding training criteria, which are weighted linear combinations of the equations.}
\label{tab:evaluated_models}
\begin{tabular}{ll}
\hline \hline
Model & Training criterion \\
\hline \hline
ACVAE & 
\refeq{modified_acvae_1}, \refeq{modified_acvae_2}, \refeq{modified_acvae_3} \\
~~+ estimated\_label & 
\refeq{modified_acvae_1}, \refeq{modified_acvae_2}, \refeq{modified_acvae_3},  
\refeq{est_l_5}, \refeq{est_l_6} \\
~~~~+ KD\_z &  
\refeq{modified_acvae_1}, \refeq{modified_acvae_2}, \refeq{modified_acvae_3},  
\refeq{est_l_5}, \refeq{est_l_6}, \refeq{KD_z}\\
~~~~+ KD\_S & 
\refeq{modified_acvae_1}, \refeq{modified_acvae_2}, \refeq{modified_acvae_3},  
\refeq{est_l_5}, \refeq{est_l_6}, \refeq{KD_true}, \refeq{KD_est_GS}\\
~~~~+ KD\_{\rm both} & 
\refeq{modified_acvae_1}, \refeq{modified_acvae_2}, \refeq{modified_acvae_3},  
\refeq{est_l_5}, \refeq{est_l_6}, \refeq{KD_z}, \refeq{KD_true}, \refeq{KD_est_GS} \\
~~+ KD\_z & 
\refeq{modified_acvae_1}, \refeq{modified_acvae_2}, \refeq{modified_acvae_3},  
\refeq{KD_z} \\
~~+ KD\_S & 
\refeq{modified_acvae_1}, \refeq{modified_acvae_2}, \refeq{modified_acvae_3}, \refeq{KD_true} \\
~~+ KD\_{\rm both} & 
\refeq{modified_acvae_1}, \refeq{modified_acvae_2}, \refeq{modified_acvae_3},  
\refeq{KD_z}, \refeq{KD_true} \\
\hline
\end{tabular}
\end{table}

\begin{table}[t]
\setlength{\tabcolsep}{4.5pt}
\centering
\caption{SDR [dB], SIR [dB], SAR [dB], PESQ, and STOI achieved by using ChimeraACVAE source model trained with different loss functions. Bold font shows the highest scores.}
\label{tab:criteria}
\begin{tabular}{llrrrrr}
\hline\hline
Scenario                             & Training criteria   & SDR & SIR & SAR & PESQ & STOI \\
\hline \hline
\multirow{8}{*}{Spk-dep}   & ACVAE  & 10.74    & 16.02    & 13.79    & 2.45     & 0.8170     \\
                                     & ~~+ estimated\_label &   13.29 &   18.87 &   15.87  &   2.64  &   0.8409    \\
                                     & ~~~~+ KD\_z     &   {\bf 15.90}  &   {\bf 22.23}  &   {\bf 17.78}  &   {\bf 2.79}  &   {\bf 0.8580}   \\ 
                               		 & ~~~~+ KD\_S\     &   13.29  &   18.75  &   16.04  &   2.66 &   0.8378   \\ 
                               		 & ~~~~+ KD\_{\rm{both}}     &   15.40  &   21.63 &   17.43 &   2.77  &   0.8565 \\ 
                               	     &  ~~+ KD\_z &   9.89 &   15.48 &   12.79 &   2.38 &   0.8114  \\
                               		 &  ~~+ KD\_S &   12.41 &   17.69 &   15.23 &   2.56 &   0.8253 \\
                               		 &  ~~+ KD\_{\rm{both}} &  8.05 &   12.96 &   11.76 &   2.24 &   0.7880 \\
\hline
\multirow{8}{*}{Spk-ind}   &   ACVAE  & 15.81    & 22.73    & 18.60    &  3.14    & 0.8855     \\
                               		 & ~~+ estimated\_label &   12.35 &   18.38 &   16.01  &   3.04    &   0.8634    \\
                                     & ~~~~+ KD\_z     &   16.89  &   24.74   &   18.79  &   3.17 &   0.8917  \\ 
                               		 & ~~~~+ KD\_S     &   15.18  &   21.95   &   17.99  &   3.12    &   0.8832   \\ 
                               		 & ~~~~+ KD\_{\rm{both}}     &   {\bf 17.04}  &   {\bf 24.87}  &  {\bf 18.85}  &   {\bf 3.19}  &   {\bf 0.8945}  \\ 
                               	    &  ~~+ KD\_z &   16.16 &   23.68 &   18.33 &   3.14 &   0.8863  \\
                               		 &  ~~+ KD\_S &   15.66 &   22.65 &   18.48 &   3.14 &   0.8893 \\
                               		 &  ~~+ KD\_{\rm{both}} &  16.07 &   23.47 &   18.39 &   3.14 &   0.8892 \\
\hline
\end{tabular}
\end{table}

\begin{table}[t!]
\setlength{\tabcolsep}{2pt}
\centering
\caption{Comparsion of SDR [dB], SIR [dB], SAR [dB], PESQ, and STOI among compact FastMVAE, FastMVAE, and FastMVAE2 with the optimal parameter settings. Bold font shows the highest scores.}
\label{tab:fmvae}
\begin{tabular}{llrrrrr}
\hline \hline
Scenario & Method     & SDR & SIR & SAR & PESQ & STOI \\
\hline\hline
\multirow{5}{*}{Spk-dep} & Compact FastMVAE w/o PoE & 7.52   & 12.71  & 10.80 & 2.29    & 0.7916    \\
& Compact FastMVAE w/ PoE  & 7.75   & 12.84   & 11.06  & 2.30 & 0.7933   \\
& FastMVAE w/o PoE [37] &  13.78   & 19.51    & 16.16    & 2.03     & 0.8465     \\
& FastMVAE w/ PoE  [37]&  13.95   & 19.54    & 16.33    & 2.66     & 0.8452     \\
& FastMVAE2 &   {\bf 15.40}  &   {\bf 21.63} &   {\bf 17.43} &   {\bf 2.77}  &   {\bf 0.8565} \\
\hline 
\multirow{5}{*}{Spk-ind} & Compact FastMVAE w/o PoE & 8.16  & 12.62 & 12.47  & 2.60 & 0.8119    \\
& Compact FastMVAE w/ PoE  & 10.55 & 17.51   & 12.66 & 2.78    &    0.8453 \\
& FastMVAE w/o PoE [37] & 10.43    & 15.41    & 15.73    & 2.73     & 0.8358     \\
& FastMVAE w/ PoE [37] & 14.41    & 21.21    & 17.35    & 3.04     & 0.8776     \\
&FastMVAE2 &   {\bf 17.04}  &   {\bf 24.87}  &  {\bf 18.85}  &   {\bf 3.19}  &   {\bf 0.8945} \\
\hline 
\end{tabular}
\end{table}

\begin{table}[t!]
\setlength{\tabcolsep}{5pt}
\centering
\caption{Comparsion of SDR [dB], SIR [dB], SAR [dB], PESQ, and STOI between FastMVAE2 and baseline methods with the optimal parameter settings. Bold font shows the highest scores.}
\label{tab:performance}
\begin{tabular}{llrrrrr}
\hline \hline
Scenario & Method     & SDR & SIR & SAR & PESQ & STOI \\
\hline\hline
\multirow{5}{*}{Spk-dep} &ILRMA  & 13.62     & 19.79    & 15.83    &  1.92    & 0.8570      \\
&IDLMA \cite{Li2020fast} & 14.15 & 21.11 & 15.59 & 1.77 & 0.8692 \\
&MVAE \cite{Li2020fast} &  {\bf 17.03}   & {\bf 23.75}    & {\bf 18.61}    & 2.24     & {\bf 0.8717}     \\
&FastMVAE  \cite{Li2020fast}      & 13.95   & 19.54    & 16.33    & 2.66     & 0.8452    \\
&FastMVAE2   &   15.40  &   21.63 &   17.43 &   {\bf 2.77}  &   0.8565  \\
\hline
\multirow{4}{*}{Spk-ind} &ILRMA  & 14.43    & 20.98    & 17.45    & 2.28     & 0.8850     \\
&MVAE \cite{Li2020fast} & {\bf 17.58}    & {\bf 25.13}    & {\bf 19.26}    & 2.65     & 0.8934     \\
&FastMVAE  \cite{Li2020fast}      &  14.41    & 21.21    & 17.35    & 3.04     & 0.8776    \\
&FastMVAE2        &   17.04  &   24.87  &  18.85  &   {\bf 3.19}  &   {\bf 0.8945}     \\
\hline
\end{tabular}
\end{table}

We first investigated the effectiveness of each training criterion proposed in \refsubsec{criterion} in training the proposed ChimeraACVAE source model.
The correspondence between the models and their training criteria are shown in \reftab{evaluated_models}.
\reftab{criteria} shows the results, which were
calculated by averaging over the entire dataset including multiple reverberation conditions.
The results show that it is effective to further exploit the reconstruction loss and classification loss of the spectrograms reconstructed with the estimated class label in the speaker-dependent scenario, where small amounts of training data were available. 
Comparing the models trained without KD (1st and 2nd rows) with that trained with KD (3th to 5th rows), we found an improvement in SDR of about 2.6 dB in speaker-dependent situations and more than 1 dB in speaker-independent ones, which confirmed that KD can significantly improve source separation performance.
In particular, knowledge transfer of the distribution of the latent variable $\Mat{z}$ was effective in stabilizing the inference accuracy even for unseen speakers. 
A further improvement was achieved in the speaker-independent setting by transferring knowledge between distributions of generated complex spectrograms, but no improvement was seen in the speaker-dependent setting.
\par
In \reftab{fmvae}, we show a comparison of source separation performance between the FastMVAE and FastMVAE2 methods. 
To demonstrate the effectiveness of the proposed training criterion, we trained an ACVAE with the architecture that respectively replaces BN and GLU with LN and SiLU, which is referred to as ``compact FastMVAE".
The results of FastMVAE and compact FastMVAE indicate that the replacement of the normalization method and nonlinear activation did not lead to an improvement of the source separation performance. 
Therefore, the performance improvement by FastMVAE2 can be attributed mainly to the proposed training criterion.
The FastMVAE2 method obtained the highest scores in terms of all the criteria. Particularly, FastMVAE2 achieved an SDR improvement of about 6.6 and 2.6 dB from the FastMVAE without and with PoE, respectively. These results indicated that the ChimeraACVAE source model had good generalization to unseen data, which made the FastMVAE2 method able to handle speaker-independent scenario without the heuristic inference method.
\reftab{performance} shows the average scores achieved by each method with their optimal parameter settings. 
The proposed method significantly outperformed ILRMA and the FastMVAE method, and narrowed the performance gap with the MVAE method. 

\subsection{Comparison of computational time in situations with more sources and channels}

\begin{figure}[t]
\centering
\includegraphics[width=\linewidth]{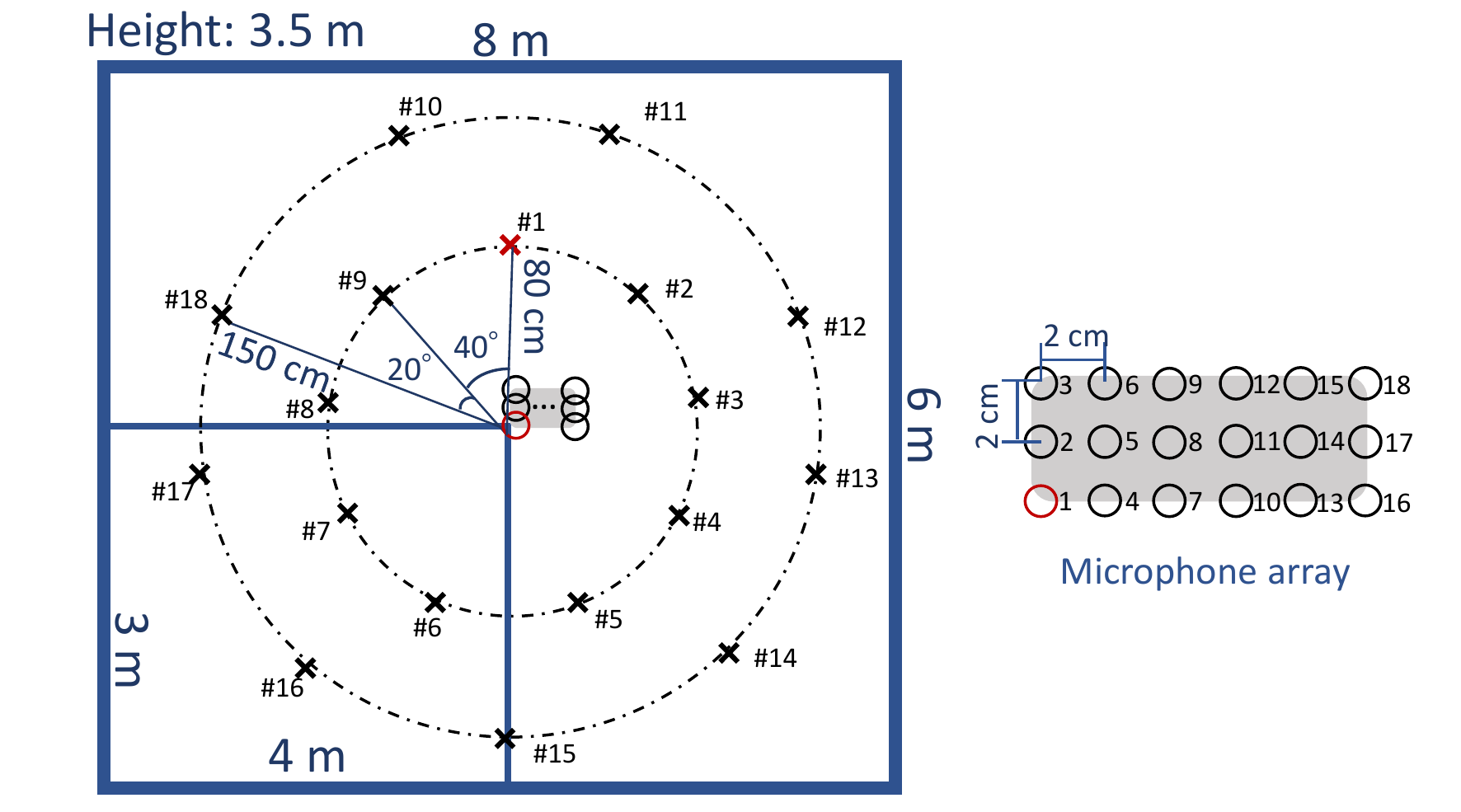} 
\caption{Configuration of sources and microphone array, where red points represent the first microphone and source.}
\label{fig:config}
\end{figure}

\begin{table}[t!]
\centering
\caption{Lengths [sec] of mixture signals in each case.}
\label{tab:length}
\begin{tabular}{crrr}
\hline\hline
Number of sources    & Minimum   & Maximum     & Average       \\
\hline\hline
2		&  5.70     &   13.86    &  8.56      \\
3        & 8.71      &   13.68    &  11.47     \\
6        & 9.04      &   16.23   &  12.76       \\
9       & 9.49   & 16.33   &  12.60    \\
12      & 10.48 & 15.32 & 12.77 \\
15      & 11.75 & 14.71 & 13.12 \\
18      & 11.43 & 15.83 & 13.51 \\
\hline
\end{tabular}
\end{table}

In this subsection, we investigate the computational time of each method. 
We conducted speaker-independent experiments with more sources and channels, and compared the computation time of each method for each update iteration and overall processing time.
\par
As in the above speaker-independent experiment,  
the simulated RIRs in the \{2, 3, 6, 9, 12, 15, 18\}-channel cases were generated using the image method \cite{Allen1979image} with the reflection coefficient of the walls set at 0.20. 
The details of the room configuration and microphone array are shown in \reffig{config}. 
In each case, more sound sources and microphones were added and placed in the order of increasing numbers.
Speech utterances were randomly selected from the WSJ0 folders $\tt si\_dt\_05$ and $\tt si\_et\_05$. 
We generated 10 samples for each case. The minimum, maximum, and average lengths of the mixture signals are shown in \reftab{length}.
The average SDR of the generated mixture signals for each case is shown in the first row of \reftab{performance_more}. 
All algorithms were processed using an Intel(R) Xeon(R) Gold 6130 CPU @ 2.10GHz and a Tesla V100 GPU.
Other experimental settings were the same as those in the above speaker-independent experiment.

\begin{figure}[t]
\centering
\includegraphics[width=\linewidth]{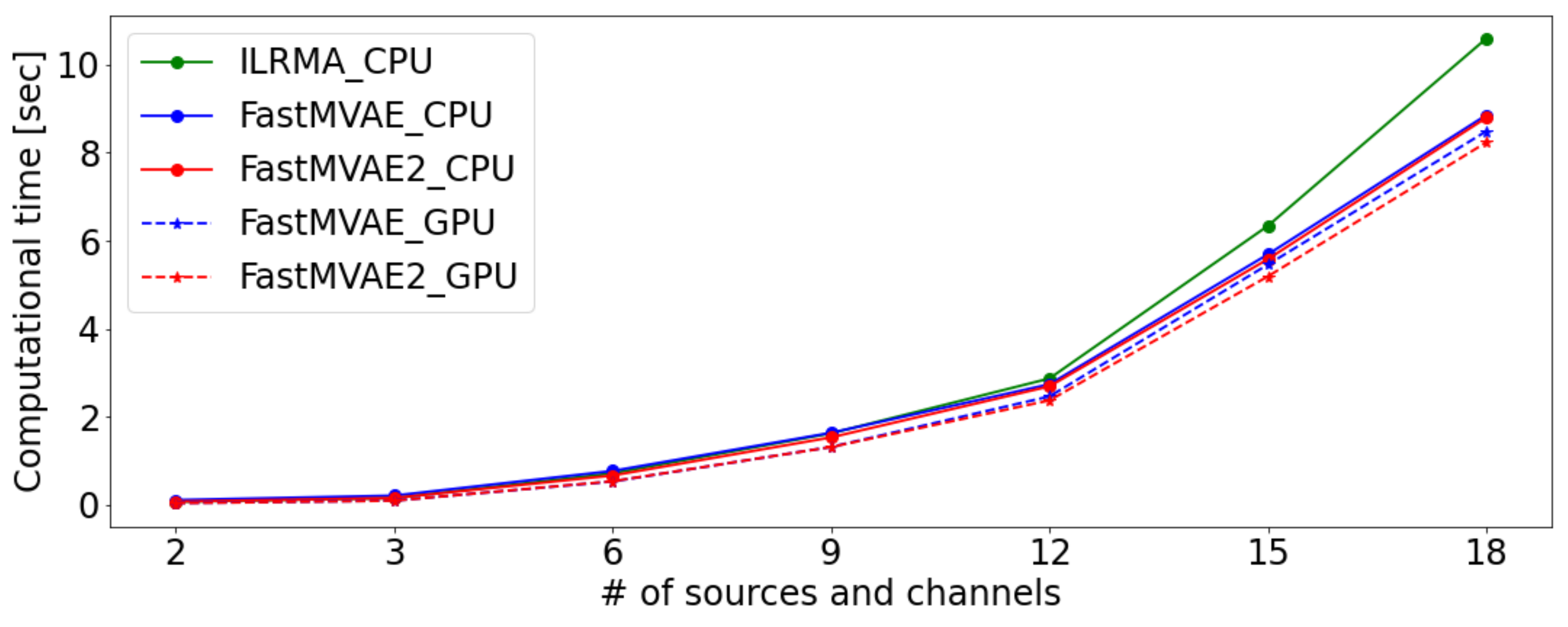} 
\includegraphics[width=\linewidth]{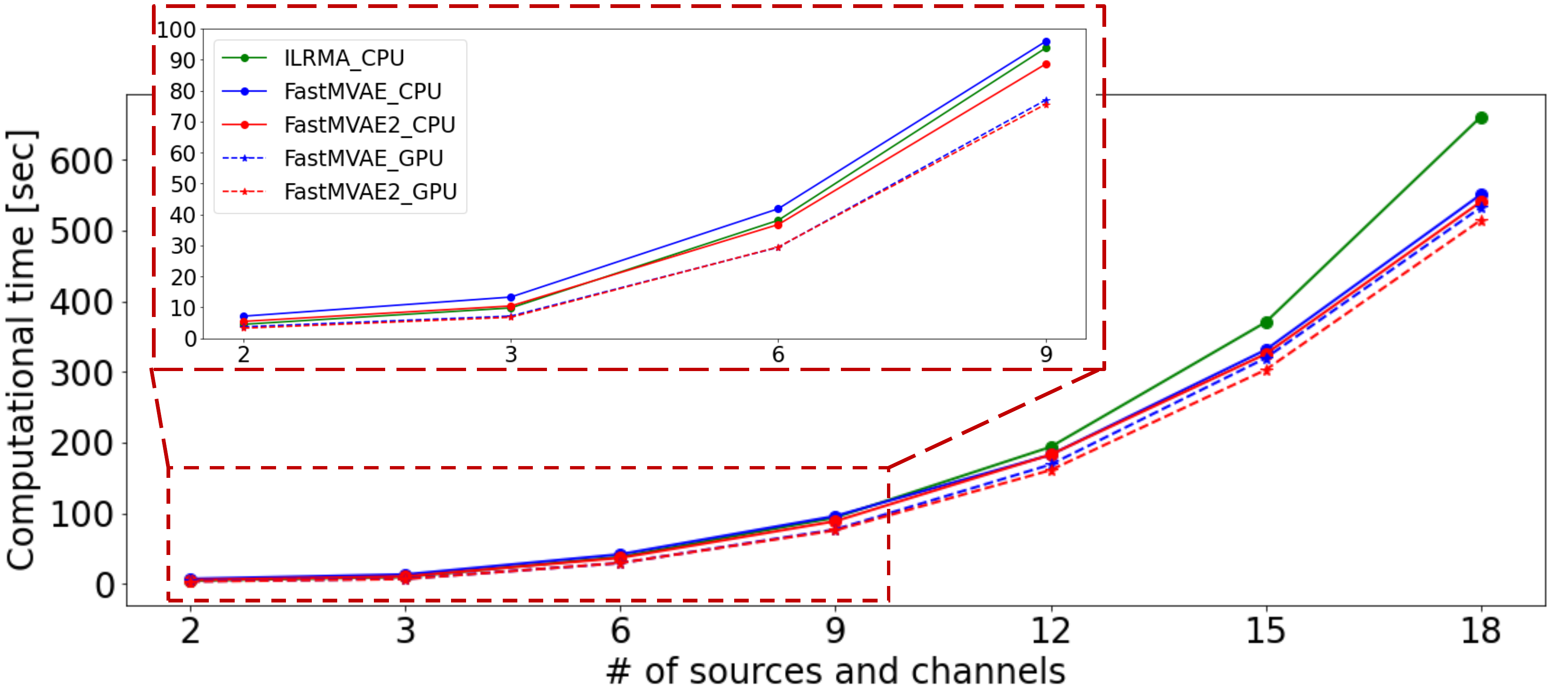} 
\caption{Average inference time [sec] of each iteration (upper) and overall processing (bottom).}
\label{fig:runtime}
\end{figure}

\begin{table}[t!]
\centering
\caption{Average inference time [sec] of MVAE.}
\setlength{\tabcolsep}{2.5pt}
\label{tab:runtime}
\begin{tabular}{lrrrrrrr}
\hline\hline
\multirow{2}{*}{Type}    & \multicolumn{7}{c}{Number of sources and channels} \\
     & \multicolumn{1}{c}{2}  & \multicolumn{1}{c}{3}     & \multicolumn{1}{c}{6} & \multicolumn{1}{c}{9}  & \multicolumn{1}{c}{12}     & \multicolumn{1}{c}{15} & \multicolumn{1}{c}{18}      \\
\hline\hline
Each iteration		&  0.70     &   1.05    &  2.65 & 4.36  & 9.24 & 10.43 & 14.03  \\
Overall processing        & 43.72      &   65.11    &  155.77 &  266.80 & 478.08 & 583.02 & 872.83  \\
\hline
\end{tabular}
\end{table}

\begin{table*}[t!]
\centering
\setlength{\tabcolsep}{8pt}
\caption{Comparison of SDR [dB] between FastMVAE2 and baseline methods with the optimal parameter settings in situations with different numbers of sources and channels. Values in parentheses indicate the improvement over unprocessed. Bold font shows the highest scores.}
\label{tab:performance_more}
\begin{tabular}{lrrrrrrr}
\hline\hline
\multirow{2}{*}{Method} & \multicolumn{7}{c}{Number of sources and channels}\\
                        & \multicolumn{1}{c}{2}  & \multicolumn{1}{c}{3}     & \multicolumn{1}{c}{6} & \multicolumn{1}{c}{9} & \multicolumn{1}{c}{12} & \multicolumn{1}{c}{15} & \multicolumn{1}{c}{18} \\
\hline\hline
Unprocessed		        & \multicolumn{1}{c}{0.09}                & \multicolumn{1}{c}{-3.92}                  & \multicolumn{1}{c}{-8.13}                 & \multicolumn{1}{c}{-10.45}            & \multicolumn{1}{c}{-12.15}            & \multicolumn{1}{c}{-13.03}        & \multicolumn{1}{c}{-13.86}      \\
ILRMA               & 20.89 (20.80)         & 23.04 (26.96)         & 7.54 (15.67)          & 1.61 (12.06)      & -0.11 (12.04)     & -3.79 (9.24) 
        & -5.92 (7.94) \\
MVAE                & 26.63 (26.54)         & {\bf 25.17 (29.09)}   & {\bf 11.32 (19.45)}   & {\bf 9.26 (19.71)}& {\bf 7.34 (19.49)}& {\bf 5.00 (18.03)}& {\bf 2.58 (16.34)} \\
FastMVAE w/o PoE    & 15.77 (15.68)         & 7.59 (11.51)          & 3.32 (11.45)          & 4.23 (14.68)      & 0.69 (12.84)      & -0.06 (12.97)         &  -1.96 (11.90)      \\
FastMVAE2           & {\bf 28.58 (28.49)}   & 21.50 (25.41)         & 6.53 (14.66)          & 5.77 (16.23)      & 4.04 (16.19)      & 2.90 (15.94)
        & 0.22 (14.08) \\
\hline
\end{tabular}
\end{table*}

\begin{figure*}[t]
\centering
\includegraphics[width=\linewidth]{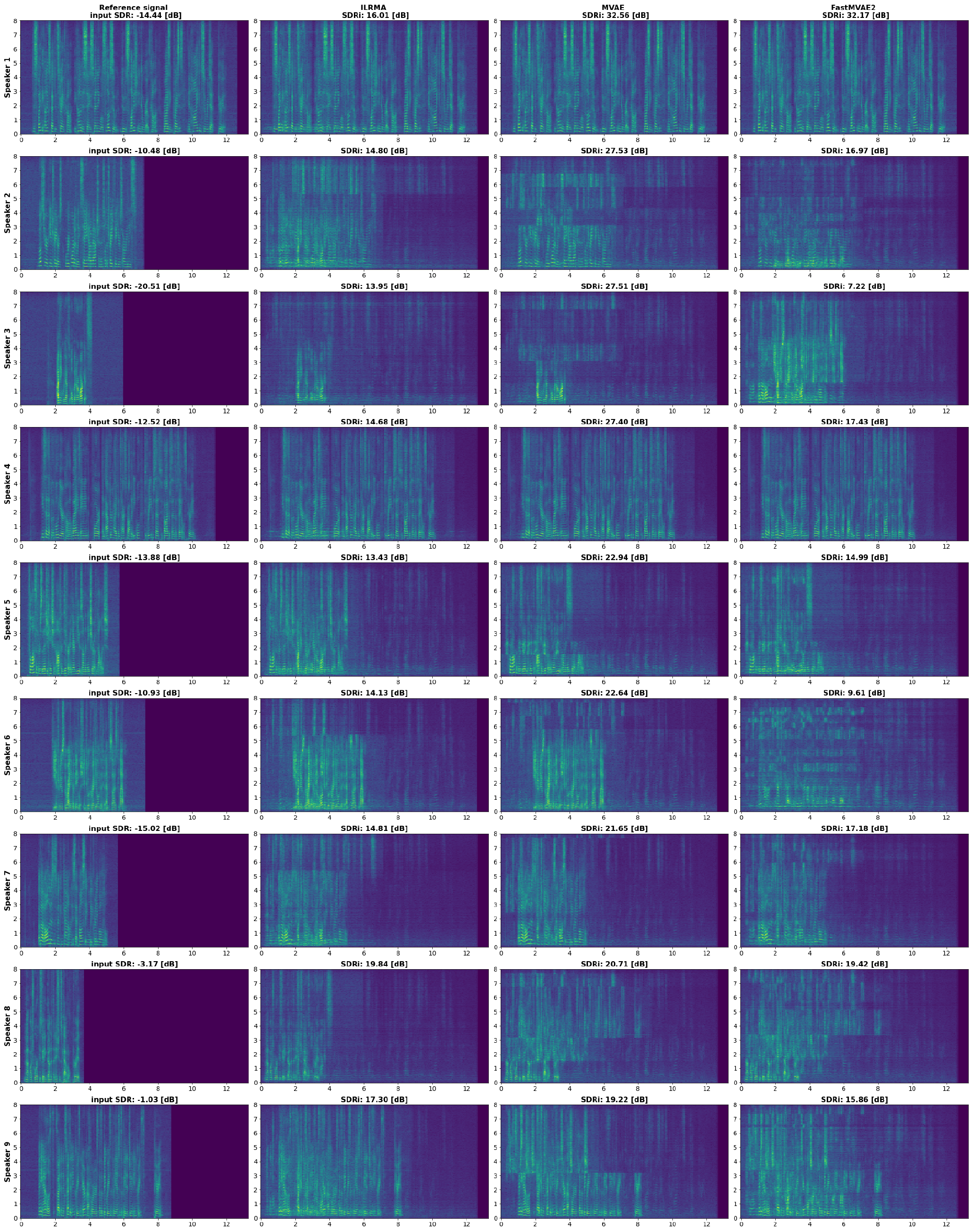} 
\caption{Magnitude spectrograms of ground truth signals (first column) and separated signals obtained by ILRMA (second column), MVAE (third column), and FastMVAE2 (fourth column) in a nine-source case. SDR of input mixture signal with respect to each speaker is shown in the top of figures in first column and SDR improvement achieved by each method is shown in the top of each figure in second to fourth. The x and y axes of each figure denote time [sec] and frequency [kHz], respectively. Audio samples are available at http://www.kecl.ntt.co.jp/people/kameoka.hirokazu/Demos/mvae-ss/index.html.}
\label{fig:spec}
\end{figure*}

\par
The inference times of ILRMA, FastMVAE, and FastMVAE2 are shown in \reffig{runtime}, and those of MVAE are shown in \reftab{runtime} as a reference. 
The fast algorithms performed extremely fast by using a GPU. 
Comparing the computation times in the CPU, we found that the FastMVAE2 method achieved runtimes comparable to ILRMA in the 2-source and 3-source cases, and faster than ILRMA in cases with more than 3 sources. 
This indicates that the proposed method is more efficient in situations with a large number of sources and microphones.
The average SDR scores obtained by each method are shown in \reftab{performance_more}.
The proposed FastMVAE2 outperformed ILRMA and the FastMVAE without PoE, and even outperformed the MVAE method in the 2-source case, demonstrating the effectiveness of the proposed ChimeraACVAE source model.
Note that although the performance of ILRMA was superior to the proposed method in the cases of 3 and 6 sources, this might change with different initialization of the basis and activation matrices of the NMF. On the other hand, the performance of the proposed method is independent of the initialization. 
We show an example of the magnitude spectrograms of separated signals obtained by ILRMA, MVAE, and FastMVAE2 with their corresponding ground truth signals in \reffig{spec}\footnote{Audio samples are available at  http://www.kecl.ntt.co.jp/people/\\kameoka.hirokazu/Demos/mvae-ss/index.html}. We found that although the MVAE and FastMVAE2 methods suffered from the phenomenon called block permutation \cite{Liang2012overcoming,Mitsui2018vectorwise}, in which the permutations in different frequency blocks are inconsistent, the deep generative model-based source models improved the estimation accuracy in the low-frequency band (0-2 kHz), which resulted in a more remarkable SDR improvement compared with ILRMA. 

\subsection{Spatialized-WSJ0-2mix benchmark}
\label{subsec:spatialized-wsj0-2mix}
In this subsection, we evaluate the proposed FastMVAE2 in the spatialized WSJ0-2mix benchmark\cite{Wang2018multi}, which is widely used for evaluating the recent DNN-based methods.
There are 20,000 ($\sim$30h), 5,000 ($\sim$10h), and 3000 ($\sim$5h) utterances in the training, validation, and test sets. 
The training and validation mixtures were generated from data in $\tt si\_tr\_s$ folder and the test mixtures were generated from data in $\tt si\_dt\_05$ and $\tt si\_et\_05$ folders. 
Therefore, the speaker-independent settings mentioned in the above experiments are still valid. 
RIR used for every utterance was simulated with a random configuration, including room characteristics, speaker locations, and microphone geometry. $RT_{60}$ for the reverberant case was randomly selected from 200 to 600 ms.
\par
We compared FastMVAE2 with (1) oracle ideal binary mask (IBM), (2) oracle ideal ratio mask (IRM), (3) oracle mask-based minimum variance distortionless response (MVDR) beamformer \cite{Trees2004optimum}, (4) oracle signal-based MVDR, 5) time-domain audio separation network (TasNet) \cite{Luo2019conv}, (6) multichannel TasNet \cite{Ochiai2020beam}, and (7) Beam-TasNet \cite{Ochiai2020beam}.
The oracle IBM and IRM were obtained using the first channel of the spatialized clean sources and applied to the first channel of observed mixture signals.
The difference between the oracle mask-based and signal-based MVDR was the signal used for computing spatial covariance matrices, where the former used the multichannel IRM of each source and the later directly used the clean reverberant speech of each source. We investigated window lengths of 128 ms and 512 ms. Settings for TasNet, multichannel TasNet, and Beam-TasNet are available in \cite{Ochiai2020beam} \footnote{We would like to appreciate Dr. Tsubasa Ochiai from NTT Communication Science Laboratories for providing us with the test dataset and evaluation script so that we could compare our methods with results reported in \cite{Ochiai2020beam}.}. One important factor here is the window length. Beam-TasNet used a length of 512 ms to meet the instantaneous mixture model for reverberant signals, while the proposed method used that of 128 ms since the motivation of the FastMVAE2 is to bridge the high performance of MVAE and real-time applications with low latency.
\par
We first show the results of the spatialized anechoic WSJ0-2mix dataset in \reftab{spatialized_ane}.
With the anechoic setup, beamforming algorithms achieved higher performance than mask-based methods. 
From these results, we confirmed that the MVAE and proposed FastMVAE2 achieved even better performance than the oracle mask-based MVDR beamformer, indicating the effectiveness of these two methods when the instantaneous mixture model assumption is satisfied.
Next, we show the results of the spatialized reverberant WSJ0-2mix dataset in \reftab{spatialized_reverb}. The performance of the MVAE and FastMVAE2 degraded significantly due to reverberations. The main reason was the instantaneous mixture model assumption, which was not satisfied anymore with heavy reverberation and short window length. We found that even the performance of oracle MVDRs degraded significantly when the window length became shorter. Two promising approaches can be considered to deal with this problem, including using longer window length and performing separation along with dereverberation \cite{Yoshioka2011blind,Kagami2018joint,Inoue2019joint}. 
It is straightforward that longer window length helps deal with heavy reverberant conditions, which has also been confirmed from the results of oracle MVDRs with longer window length and Beam-TasNet. However, a longer window length is undesirable and should be avoided in real-time applications because it increases algorithmic latency.
Therefore, we consider the second approach, performing separation and dereverberation simultaneously, as one direction of our future works to overcome this limitation of the FastMVAE2 method.

\begin{table}[t!]
\setlength{\tabcolsep}{4pt}
\centering
\caption{Comparison of SDR [dB] for spatialized anechoic WSJ0-2mix dataset. ``1ch" and ``2ch" indicate the number of channels used for processing.}
\label{tab:spatialized_ane}
\begin{tabular}{lcr}
\hline \hline
Method  &  window length [ms] & SDR [dB] \\
\hline\hline
Mixture & --- & -0.4\\
\hline
Oracle IBM (1ch)&  128 & 13.66 \\
Oracle IRM (1ch) & 128 & 13.55 \\
Oracle mask-based MVDR  (2ch) & 128 & 23.26   \\
Oracle signal-based MVDR (2ch) & 128 & 39.68   \\
Oracle mask-based MVDR  (2ch) & 512 & 15.66   \\
Oracle signal-based MVDR (2ch) & 512 & 48.02   \\
\hline
MVAE (2ch) & 128 & 28.49 \\ 
FastMVAE2 (2ch) & 128 & 31.31 \\  
\hline
\end{tabular}
\end{table}

\begin{table}[t!]
\centering
\caption{Comparison of SDR [dB] for spatialized reverberant WSJ0-2mix dataset. ``1ch" and ``2ch" indicate the number of channels used for processing.}
\label{tab:spatialized_reverb}
\begin{tabular}{lcr}
\hline \hline
Method     &  window length [ms] & SDR [dB]  \\
\hline\hline
Mixture & --- & 0.1  \\
\hline
Oracle IBM (1ch) & 128 & 13.41 \\
Oracle IRM (1ch) & 128 & 13.29 \\
Oracle mask-based MVDR (2ch) & 128 & 8.16    \\
Oracle signal-based MVDR (2ch)  & 128 & 8.14   \\
Oracle mask-based MVDR (2ch) & 512 & 11.95    \\
Oracle signal-based MVDR (2ch)  & 512 & 16.32   \\
\hline
TasNet (1ch) \cite{Ochiai2020beam} & --- & 11.3  \\
Multichannel TasNet (2ch) \cite{Ochiai2020beam} & --- &  12.7   \\
Beam-TasNet (1ch) \cite{Ochiai2020beam} & 512 & 12.9 \\
Beam-TasNet (2ch) \cite{Ochiai2020beam} & 512 & 13.8 \\
\hline
MVAE (2ch) & 128 & 5.35 \\ 
FastMVAE2 (2ch) & 128 & 6.02 \\ 
\hline
\end{tabular}
\end{table}

\section{Conclusion}
\label{sec:conclusion}
In this paper, we proposed an improved ACVAE source model named ``ChimeraACVAE" source model for the fast algorithm of the MVAE method, which we call ``FastMVAE2". 
ChimeraACVAE is a more compact source model that consists of a unified encoder and classifier network and a decoder, which are composed of fully convolutional layers with layer normalization and an SiLU activation function. The KD framework was applied to train the ChimeraACVAE source model to improve the generalization capability to unseen data.
The experimental results demonstrated that the FastMVAE2 method achieved significant performance improvement in both speaker-dependent and speaker-independent multispeaker separation tasks, approaching the performance that of the MVAE method.
Moreover, the proposed method significantly reduced the model size and improved the computational efficiency, which achieved computational time comparable to ILRMA in cases of two and three sources and lower computational time than ILRMA in cases of more sources.

\begin{IEEEbiography}{Li Li}
received the B.E. degree from Shanghai University of Finance and Economics, China, in 2014, and the M.S. and Ph.D. degrees from the University of Tsukuba, Japan, in 2018 and 2021, respectively. She is currently a postdoctoral researcher with Nagoya University and an adjunct researcher with  NTT Communication Science Laboratories, Nippon Telegraph and Telephone Corporation. Her research interests include audio and speech signal processing, source separation, and machine learning. She received the Student Presentation Award from the Acoustical Society of Japan and the Signal Processing Society Japan Student Conference Paper Award.
\end{IEEEbiography}

\begin{IEEEbiography}{Hirokazu Kameoka}
received the B.E., M.S. and Ph.D. degrees from the University of Tokyo, Japan, in 2002, 2004, and 2007, respectively. He is currently a Senior Distinguished Researcher at NTT Communication Science Laboratories, Nippon Telegraph and Telephone Corporation and an Adjunct Associate Professor at the National Institute of Informatics. From 2011 to 2016, he was an Adjunct Associate Professor at the University of Tokyo. His research interests include audio, speech, and music signal processing and machine learning. He has been an associate editor of the IEEE/ACM Transactions on Audio, Speech, and Language Processing since 2015, a Member of IEEE Audio and Acoustic Signal Processing Technical Committee since 2017, and a Member of IEEE Machine Learning for Signal Processing Technical Committee since 2019. He has received 17 awards, including the IEEE Signal Processing Society 2008 SPS Young Author Best Paper Award. He is the author or co-author of about 150 articles in journal papers and peer-reviewed conference proceedings.
\end{IEEEbiography}

\begin{IEEEbiography}{Shoji Makino}
received his B.E., M.E., and Ph.D. degrees from Tohoku University, Japan, in 1979, 1981, and 1993, respectively. He joined NTT in 1981. He is now a Professor at the University of Tsukuba. His research interests include adaptive filtering technologies, the realization of acoustic echo cancellation, blind source separation of convolutive mixtures of speech, and acoustic signal processing for speech and audio applications.
\par
He received the ICA Unsupervised Learning Pioneer Award in 2006, the IEEE MLSP Competition Award in 2007, the IEEE SPS Best Paper Award in 2014, the Achievement Award for Science and Technology by the Minister of Education, Culture, Sports, Science and Technology in 2015, the Hoko Award of the Hattori Hokokai Foundation in 2018, the Outstanding Contribution Award of the IEICE in 2018, the Technical Achievement Award of the IEICE in 2017 and 1997, the Outstanding Technological Development Award of the ASJ in 1995, and 8 Best Paper Awards. He is the author or co-author of more than 200 articles in journals and conference proceedings and is responsible for more than 150 patents. He was a Keynote Speaker at ICA2007 and a Tutorial Speaker at EMBC2013, Interspeech2011 and ICASSP2007.
\par
He has served on IEEE SPS Board of Governors (2018-20), Technical Directions Board (2013-14), Awards Board (2006-08), Conference Board (2002-04), and Fellow Evaluation Committee (2018-20). He was a member of the IEEE Jack S. Kilby Signal Processing Medal Committee (2015-18) and the James L. Flanagan Speech \& Audio Processing Award Committee (2008-11). He was an Associate Editor of the IEEE Transactions on Speech and Audio Processing (2002-05) and an Associate Editor of the EURASIP Journal on Advances in Signal Processing (2005-12). He was a Guest Editor of the Special Issue of the IEEE Signal Processing Magazine (2013-14). He was the Chair of SPS Audio and Acoustic Signal Processing Technical Committee (2013-14) and the Chair of the Blind Signal Processing Technical Committee of the IEEE Circuits and Systems Society (2009-10). He was the General Chair of IWAENC 2018, WASPAA2007, IWAENC2003, the Organizing Chair of ICA2003, and is the designated Plenary Chair of ICASSP2012.
Dr. Makino is an IEEE SPS Distinguished Lecturer (2009-10), an IEEE Fellow, an IEICE Fellow, a Board member of the ASJ, and a member of EURASIP.
\end{IEEEbiography}
\end{document}